\documentclass[rpreprint,3p,11pt,times,sort&compress, colorlinks,linkcolor=blue, citecolor=blue]{elsarticle}
\usepackage{amssymb}
\usepackage{amsmath,mathrsfs,lineno}
\usepackage{stfloats}
\usepackage{bm}
\usepackage{graphicx}
\usepackage{booktabs}
\usepackage{float}
\usepackage{caption}
\usepackage{sectsty}
\usepackage{threeparttable}
\usepackage{gensymb}

\usepackage[toc,page]{appendix}
\usepackage[colorlinks,linkcolor=blue,anchorcolor=green,citecolor=blue]{hyperref}

\journal{***} 
\renewcommand*{\today}{February 10, 2023}

\sectionfont{\fontsize{11}{11}\selectfont}
\begin{document}
\begin{frontmatter}

\title{Modeling and simulation of sintering process across scales}

\author[rvt]{Min Yi\corref{cor1}}
\ead{yimin@nuaa.edu.cn}

\author[rvt]{Wenxuan Wang}
\author[rvt]{Ming Xue}
\author[rvt,loc]{Qihua Gong\corref{cor1}}
\ead{gongqihua@nuaa.edu.cn}

\author[tud]{Bai-Xiang Xu}

\cortext[cor1]{Corresponding author at: College of Aerospace Engineering, NUAA, Nanjing 210016, China }



\address[rvt]{National Key Lab of Mechanics and Control of Aerospace Structures \& Key Lab for Intelligent Nano Materials and Devices of Ministry of Education \& Institute for Frontier Science \& College of Aerospace Engineering, Nanjing University of Aeronautics and Astronautics (NUAA), Nanjing 210016, China}
\address[loc]{MIIT Key Lab of Aerospace Information Materials and Physics \& College of Physics, Nanjing University of Aeronautics and Astronautics (NUAA), Nanjing 211106, China}
\address[tud]{Institute of Materials Science, Technische Universit\"at Darmstadt, Darmstadt 64287, Germany}


\begin{abstract}
Sintering, as a thermal process at elevated temperature below the melting point, is widely used to bond contacting particles into engineering products such as ceramics, metals, polymers, and cemented carbides. Modelling and simulation as important complement to experiments are essential for understanding the sintering mechanisms and for the optimization and design of sintering process. We share in this article a state-to-the-art review on the major methods and models for the simulation of sintering process at various length scales. It starts with molecular dynamics simulations deciphering atomistic diffusion process, and then moves to microstructure-level approaches such as discrete element method, Monte--Carlo method, and phase-field models, which can reveal subtle mechanisms like grain coalescence, grain rotation, densification, grain coarsening, etc. Phenomenological/empirical models on the macroscopic scales for estimating densification, porosity and average grain size are also summarized. The features, merits, drawbacks, and applicability of these models and simulation technologies are expounded. In particular, the latest progress on the modelling and simulation of selective and direct-metal laser sintering based additive manufacturing is also reviewed. Finally, a summary and concluding remarks on the challenges and opportunities are given for the modelling and simulations of sintering process.
\end{abstract}


\end{frontmatter}

\section{Introduction}
Sintering as a thermal process for consolidating powders or particles into the densified components has been critical and essential for the manufacturing industry of the modern society~\cite{German1996Sintering}.
For example, a sintering process is the first choice for producing different kinds of hard metal or ceramic components. In the area of powder metallurgy, sintering is popularly applied to produce the expected products from a porous assembly of powders. The design and optimization of sintering process play a critical role in the manufacturing of sintering-based products, which depend on the detailed understanding of the physical processes and the quantitative models at different scales. In this aspect, modelling and simulation of sintering process across scales are indispensable complements to experimental trials and errors, especially for the overall goal of producing a condensed body (from rather friable green bodies) with controlled porosity, microstructure, grain size, grain distribution, etc.

In general, sintering is driven by energy minimization through eliminating/decreasing surface area and grain boundaries. It is, depending on the process conditions, often also complicated by additional aspects such as mass/solute transport, temperature, pressure, external fields and environment~\cite{WALKER1955Mechanism}. Modelling and simulation techniques should be capable of describing and predicting the spatial and temporal microstructure evolution and its dependence on the processing and material parameters such as time, temperature, density, pressure, particle size, etc. It is generally difficult to achieve qualified products by sintering within one-time trial. Compared to experimental efforts, modeling and simulations are much more efficient in terms of cost and time period. Moreover, most simulations are by nature \textit{in-situ}, resolving transient behavior during sintering process.

The development of modelling and simulation techniques for sintering process has begun in the 1940s and emerged as an important subdiscipline of sintering research, contributing to the formation of sintering science beyond the experience-based technology~\cite{Bordia2009Advances,Kang2012Advances,Bordia2017Sintering}. The emphasis of modelling and simulation of sintering is to predict the microstructure evolution and the macroscopic quantities by using theory, modelling, computation, and analysis in different time and space scales. Especially in the recent decades, due to the availability of high-performance clusters and numerical software tools for different scales, computation studies on the complex sintering process of large systems are becoming feasible.

\begin{figure*}[!b]
\centering
\includegraphics[width=16cm]{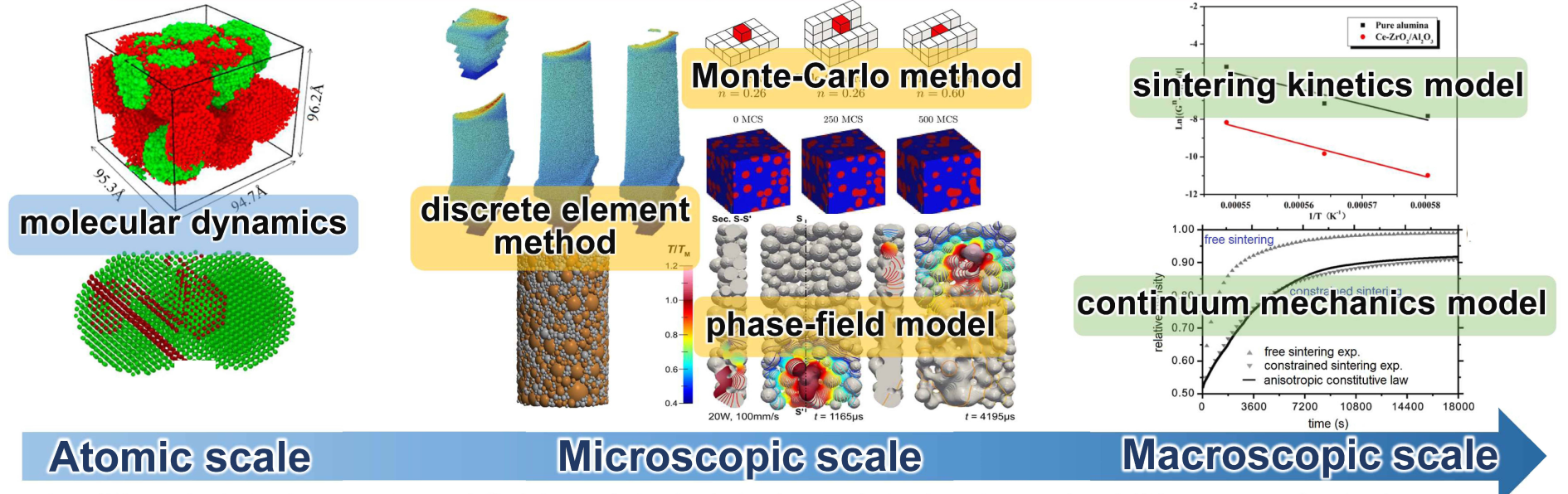}
\caption{Modeling and simulation of sintering process at various scales by different methodologies.}
\label{fig01}
\end{figure*}

Several recent review papers have provided an overview on the up-to-date development of sintering science and technology, including the continuum mechanical models and microstructure-level models~\cite{Bordia2017Sintering,Raether2019Simulation}. However, the latest progress on the application of molecular dynamics and phase-field models to sintering is not surveyed.
In addition, up to now sintering involves the traditional isothermal technique and the additive manufacturing technology. Selective laser sintering (SLS) has been a popular and promising additive manufacturing technology realized by the sintering mechanism~\cite{Olakanmi2015Areview,Sing2017Direct}. 
In contrast to the traditional isothermal sintering where a constant temperature is assumed for a sufficiently long period, the sintering technique implemented by SLS additive manufacturing possesses the characteristics of extremely non-uniform temperature distribution and high temperature gradient, which pose challenges to the modelling and simulation techniques that are created for the traditional sintering. Therefore in this review, we will exhaustively survey the recent progress on the across-scale modelling and simulation of sintering and cover the atomic, microstructural and macroscopic approaches. Moreover, the latest progress on the modelling and simulation of SLS is also presented.

This review is structured as follows. In Section~\ref{sec1}, the typical features of traditional sintering and SLS are briefly overviewed. As illustrated in Fig. \ref{fig01}, Section~\ref{sec2} comprehensively summarizes the state-of-art modelling and simulation methodology of sintering across different scales, covering atomic-scale method, microstructure-scale method (discrete element method, Monte--Carlo simulation, phase-field model), and macroscopic continuum model, in which the latest progress of applying these methods to SLS is also presented. Finally, Section~\ref{sec3} gives a short summary of this review, as well as an outlook for the possible future directions and chances in the community of modelling and simulation for sintering process.

\section{Features of sintering} \label{sec1}
It is a long history that human beings have utilized sintering technique to synthesize ceramics, refractory materials, high-temperature materials, etc~\cite{German1996Sintering}. In most cases, the sintered products from powders or particles are of polycrystal nature, consisting of crystal grain, glass structure, porosity, etc~\cite{Stuijts1973Synthesis, Yu2017Review}. 
The sintering process parameters have noticeable influences on the grain size, porosity size, grain boundary shape, etc., and thus determine the properties and performance of the sintered products.

Diffusion is one major mechanisms involved in sintering. The diffusion process is extremely slow in solid powders or particles at room temperature. So sintering is often operated at high temperatures that significantly accelerate diffusion. However, in order to control the desirable phases and avert the material decomposition, the operation temperature is generally below the melting temperature. In most cases, sintering has to be operated in vacuum or a special gas atmosphere. For some special systems, gas pressure is also an important factor for controlling sintering process. For instance, oxygen gas pressure has to be accurately controlled for the sintering of copper oxides based high-temperature superconductor, in order to obtain the designed structure, composition, and copper-valence distribution.

Sintering process can be divided into three stages \cite{Thummler1967}. The first stage is the growth of sintering neck at the initial state. In this stage, the neck growth obeys the exponential law as a function time. There is no apparently quick growth of the pristine particles, so the particles could be approximately regarded as isolated ones. The surface tension will make the adjacent particles contact and fix the grain boundary. The particles mildly shrink and the center of adjacent particles is slightly closer. The second stage is the densification together with grain growth. As the sintering neck continues growing, initially separated particles are gradually bonded together. Substantial shrinkage happens to form a network of pores, and grain coarsening continues. Grain boundary is extended from one pore to another pore. Once the effective density exceeds 90\% of the theoretically maximum density, lots of pores will be annihilated and densification occurs to approach the final sintering stage. The third stage is further densification, but the process is extremely slow. There are always residual pores and perfect densification with a density equalling to the theoretically maximum one is generally impossible.

Sintering has notable time-lag effect. The reaction during sintering is very slow. When the temperature is gradually increased and reaches the end point, the sintering process will not cease suddenly. The examination of sintered products cannot reflect the influence of processing parameters on the evolution of microstructure during the sintering process. Sintering is also a complex phenomenon, relating to thermodynamics, kinetics, chemistry, crystallization, phase transformation, etc. The sintering process itself is highly nonlinear, including multiphysics coupling, strong nonlinear dependence of material parameters on temperature and phase, nonlinear kinetics, etc. As for the SLS technique~\cite{Singh2016Progress},
differing from traditional sintering, the powders are not consolidated prior to heating and often partial melting occurs. In SLS, the laser-matter interaction and the extremely high temperature gradients will make sintering process more complex and nonlinear, challenging the current approaches of modelling and simulation of sintering with the consideration of the special features of SLS.

\section{Modelling and simulation methodology of sintering} \label{sec2}

Similar to other fields of materials science and technology, the modelling and simulation of sintering process should also cover different time and spatial scales~\cite{Rojek2017}. Different theoretical models and the associated numerical methods are available, each suitable for one specific scale. The integration of models and numerics that span a wide range of scales, scale bridging strategies, and their shrewd combination with experiments are the leading-edge directions. A typical application of integrated computational materials engineering (ICME) methodology to sintering is highly recommended. In the field of sintering, the atomic-scale method, microstructure-scale method, and macroscopic continuum model will be overviewed in the following.

\subsection{Atomic-scale method}
For the atomic-scale and early-stage sintering of nanoparticle systems, molecular dynamics (MD) has been widely used. 
In MD simulations, Newton’s equations of motion are solved to compute the temporal and spatial evolution of an assembly of interacting atoms within a short time (usually less than micro seconds). For such techniques, the force field or potential that describes how the atoms or molecules will interact with each other is critically important.
MD simulations favor the mechanistic atomic-scale understanding of sintering process in terms of atomic diffusion, nanoparticle rotation, atom trajectories, etc., and provide novel insights for the sintering of nanoscale particulate systems.
However, the time size and model sample that MD can handle is usually  restricted to less than 1~$\mu$s and 1~$\mu$m, respectively, owing to the relatively high computational cost of MD. More specifically, most MD simulations of sintering only contain nanoparticles with the size of several to several tens nanometers. MD simulations have demonstrated the influence of sintering temperature, material type, number and size of nanoparticles on the sintering behavior.

\begin{figure}[!b]
\centering
\includegraphics[width=8.4cm]{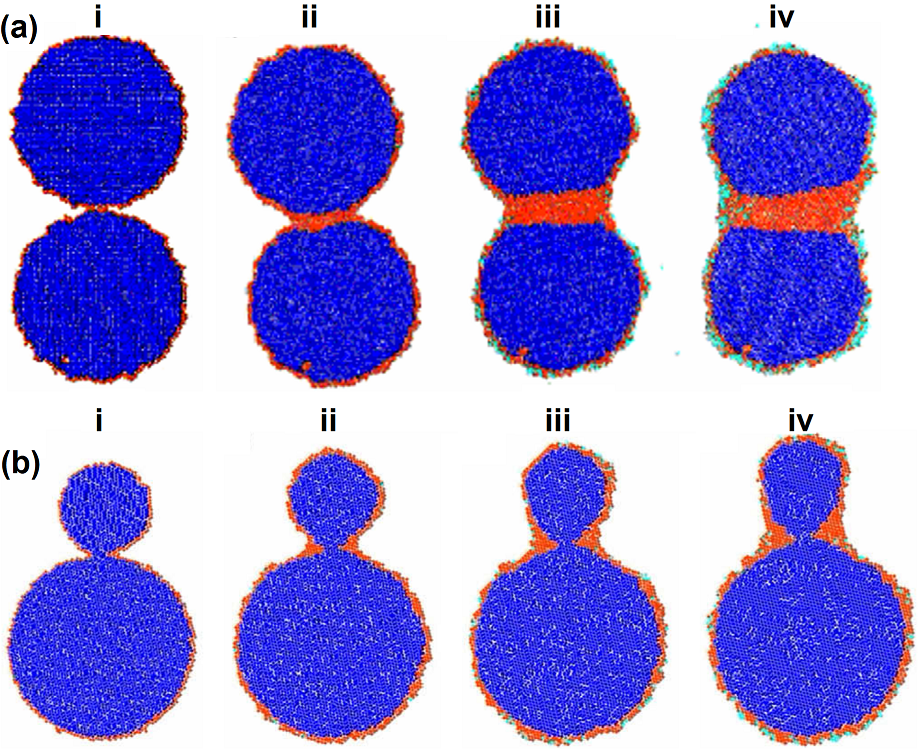}
\caption{MD simulation results of (a) two large nanoparticles sintered at 0.63$T_\text{m}$ and (b) a small and a large nanoparticle sintered at 0.46$T_\text{m}$. (i) $\tilde{t}=500$; (ii) $\tilde{t}=50000$; (iii) $\tilde{t}=200000$; (iv) $\tilde{t}=500000$. $T_\text{m}$ is the melting temperature. Reproduced with permission~\cite{Ding2009Amolecular}. Copyright 2009, Elsevier.}
\label{figMD1}
\end{figure}

For two geometrically equivalent nanoparticles composed of the same material (two-dimensional Lennard--Jones crystal) in Fig.~\ref{figMD1}(a)~\cite{Ding2009Amolecular}, under a sintering temperature of 0.63$T_\text{m}$ ($T_\text{m}$ as the melting temperature), a grain boundary forms between the two particles. At the initial stage, atoms in these two nanoparticles do not diffuse into the vacuum. Meanwhile, the surfaces contact to favor the subsequent formation of a grain boundary. The grain-boundary diffusion promotes the neck-size growth. The particles approach each other, indicating that the grain-boundary diffusion takes away atoms from the grain boundary. In contrast, for a small and a large nanoparticles sintered at 0.46$T_\text{m}$ Fig.~\ref{figMD1}(b)~\cite{Ding2009Amolecular}, there is no grain boundary formed between the two particles and the particle-particle distance keeps unchanged. Grain-boundary diffusion dose not occur and the bulk deformation is also negligible. At a lower sintering temperature of 0.46$T_\text{m}$ in Fig.~\ref{figMD1}(b), surface diffusion dominates the matter redistribution.
Similar studies are carried out for other materials system. In addition to the surface atom diffusion and neck growth, the crystal structure change is also examined~\cite{Liu2020,Zhang2016,LiuJunpeng2021Sintering,Goudeli2016Crystallinity}. For Fe$_2$O$_3$, it is found that HCP and BCC are gradually changed into amorphous structures whose proportion is around 66\%~\cite{Liu2020}.

\begin{figure}[!b]
\centering
\includegraphics[width=8.4cm]{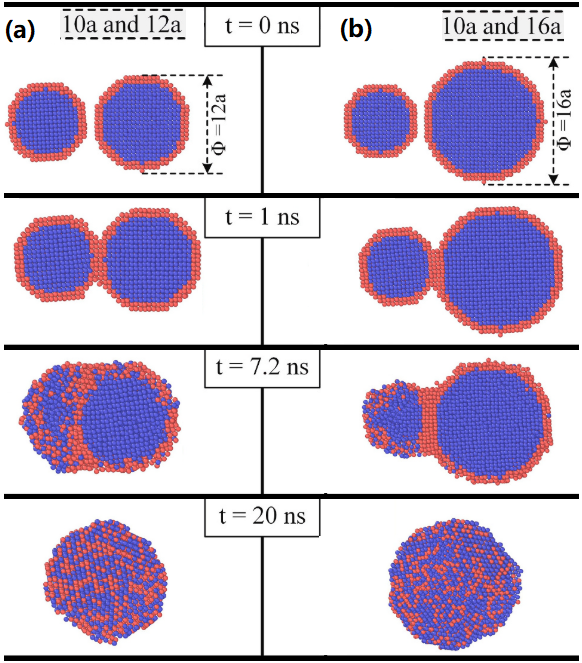}
\caption{MD simulations of sintering process of two nanosilver spheres with different size. Reproduced with permission~\cite{Zhang2021}. Copyright 2021, Elsevier.}
\label{figMD2}
\end{figure}

More MD simulation results on two nanoparticles of the same material can be found, e.g.,~\cite{Ding2009Amolecular,Zhang2021,Sementa2018Molecular}. Fig.~\ref{figMD2} shows the sintering process of two nanosilver balls of different size. The atoms in nanosilver ball gradually migrate from the outside surface to the inside body. The big nanoball seems to swallow the small one. It is revealed that as temperature increases, the sintering mechanism is changed from surface diffusion to volume diffusion. For nanosilver particles with different diameters, the sintering process is lagging. The thermal conductivity of the sintered silver nanoparticles is also examined~\cite{Zhang2021}. Similar sintering features have been observed in simulation results of other nanoparticles, e.g., Cu, Ni, TiO$_{2}$~\cite{Raut1998,Meng2019,Koparde2005,Koparde2008,Koparde2008Sintering, Zhang2020,Seong2016,Samsonov2019,Yousefi2015,cao2021}.

\begin{figure}[!b]
\centering
\includegraphics[width=8.4cm]{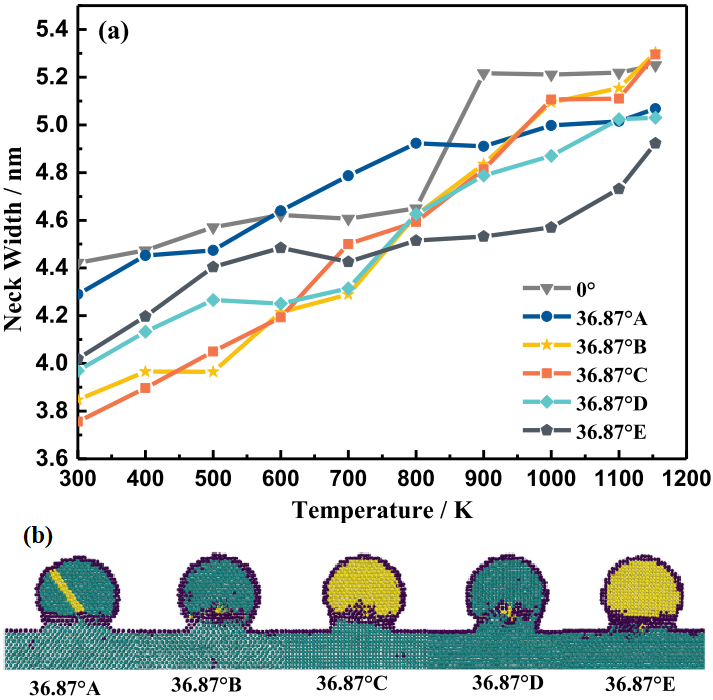}
\caption{(a) Temperature dependent neck width of Cu nanoparticle-plate with $\Sigma$5 [010] twisted GBs. (b) Atomic structure (nanoparticle diameter 4.392 nm, 200 ps, 500 K).  Reproduced with permission~\cite{Zhan2021}. Copyright 2021, Elsevier.}
\label{figMD7}
\end{figure}

Specifically, Cu nanoparticle-plate sintering system has been simulated by MD~\cite{Zhan2021}. The influence of tilt grain boundaries (GBs) ($\Sigma$5 twisted GBs) on the sintering kinetics is examined. The neck width at different temperatures is shown in Fig.~\ref{figMD7}(a) for five kinds of $\Sigma$5 [010] twisted GBs. In contrast to the non-oriented sphere-plate model, the neck growing speed of misorientation structure is almost constant during the whole heating process. From the atomic structure in Fig.~\ref{figMD7}(b), it can be seen that five different $\Sigma$5 [010] twisted GBs lead to different sintering behaviors. Particle-particle misalignment also favors an enhancement in GB diffusion~\cite{Zhan2021}.


\begin{figure}[!b]
\centering
\includegraphics[width=8.4cm]{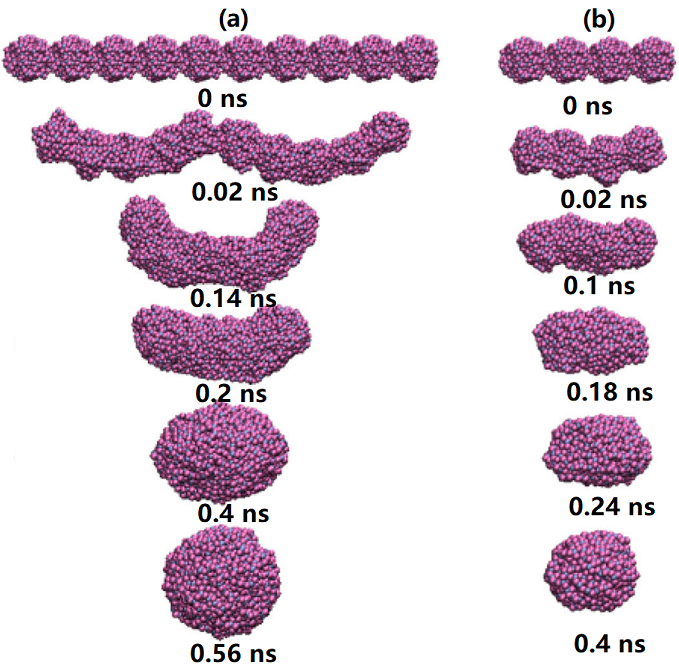}
\caption{MD simulations of sintering multiple TiO$_2$ nanoparticles at 1480~K: (a) 10 and (b) 4 nanoparticles. Reproduced with permission~\cite{Mao2015}. Copyright 2015, Springer.}
\label{figMD3}
\end{figure}

Sintering of multiple nanoparticles of the same material is explored by MD simulations despite of the high computation cost, such as TiO$_2$~\cite{Mao2015},
 Al~\cite{He2019Molecular}, Ag~\cite{Alarifi2013}, ITO~\cite{Zhou2013,Peng2013}, Pt~\cite{Suzuki2009}, Fe~\cite{Nguyen2011}, etc. Fig.~\ref{figMD3} shows the sintering process of many TiO$_2$ nanoparticles in the form chains via MD simulations. The sintering process for 10 nanoparticles in Fig.~\ref{figMD3}(a) indicates that nanoparticles at the two ends migrate toward the center at the initial stage. The quick neck growth makes the particles' original shape rapidly unrecognizable. Then a U shape forms. Subsequently, the side chains move back and a cylinder-like shape forms. At the end, for the minimization of surface energy, the particle finally becomes spherical~\cite{Mao2015}.
The sintering dynamics of 4 nanoparticles in Fig.~\ref{figMD3}(b) experiences similar behaviors, but the chain shape is similar to half of the chain with 10 nanoparticles~\cite{Mao2015}.

\begin{figure}[!b]
\centering
\includegraphics[width=8.4cm]{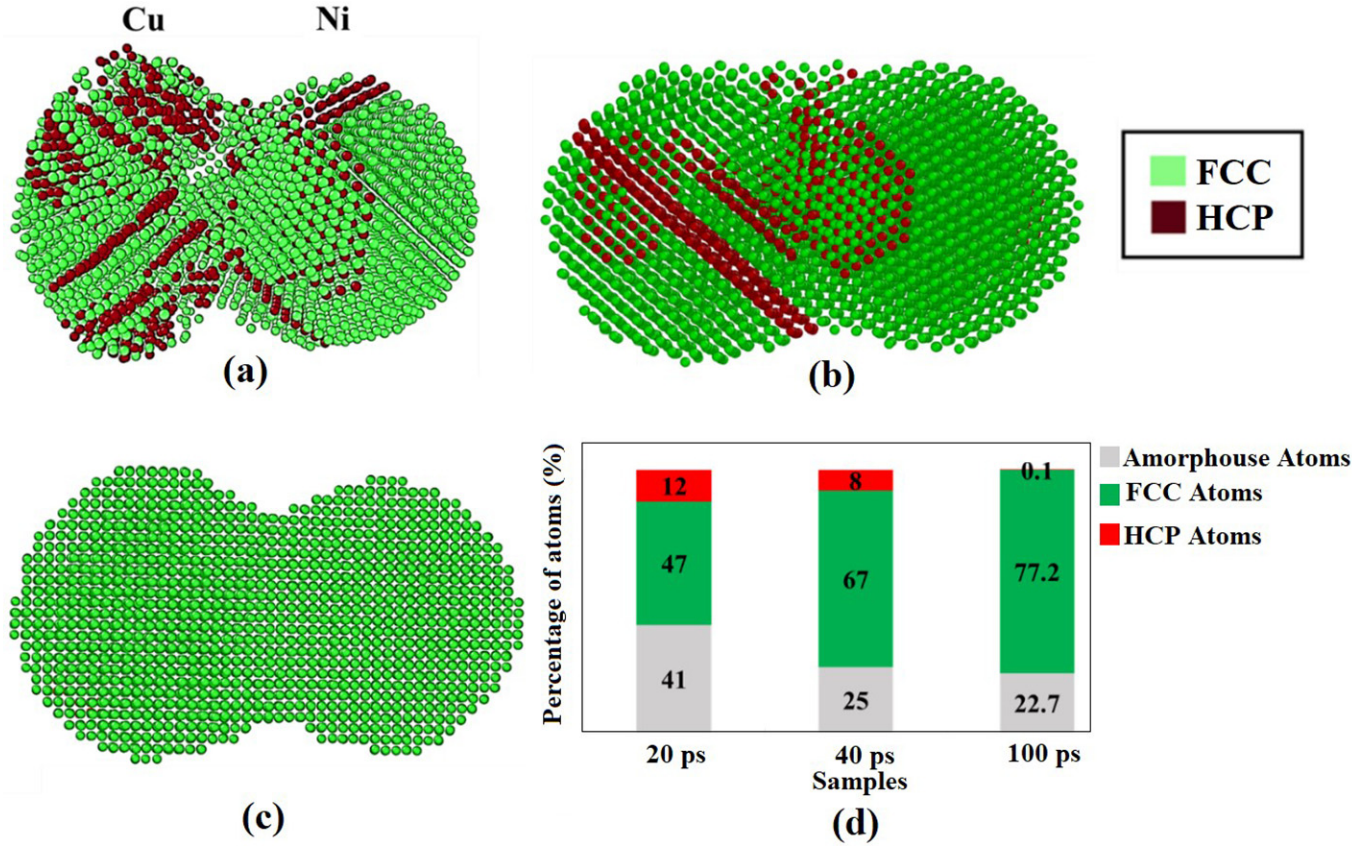}
\caption{MD simulations results on the finally achieved microstructure of Cu-Ni nanoparticles which are sintered at 1000 K for different sinering time: (a) 20 ps, (b) 40 ps, and (c) 100 ps. (d) Content of FCC, HCP, and amorphous phase. Reproduced with permission~\cite{Malti2021}. Copyright 2021, Elsevier.}
\label{figMD6}
\end{figure}

In addition, the sintering of nanoparticles with different materials provides more degrees of freedom to manipulate the microstructure and properties.
MD methods are applied to examine the nanostructure evolution throughout the sintering process of Cu and Au nanoparticles~\cite{Zhang2019,Tian2020}, with a focus on the changes in crystalline and sintering neck.
The sintering process of Al and Ni nanoparticles by MD simulations~\cite{Henz2009} shows that Al-rich compounds form at the initial stage and then the eutectic alloy rapidly forms. MD simulations of the surface evolution during the sintering of Cu-Ag nanoparticles indicate a formation of Cu-core@Ag-shell nanoparticles from the Cu/Ag alloys when Ag content is excessive. As the temperature increases, the sintered system gradually changes from the separate nanoparticles to a uniform Cu/Ag alloy~\cite{Liang2020Surface}. 
The sintering behavior of Cu-Ni nanopowders is also explored by MD simulations, with a focus on the microstructural analysis and the interfacial evolution at the atomic scale. It is found that the sintering mechanism is temperature dependent. From 600 to 1000 K, dislocation slip dominates the process. Beyond 1000 K, the thermal twinning and surface diffusion predominate. In the case of 1000 K sintering for 100 ps, the combination of surface diffusion and dislocations slip would lead to a defect-less structure~\cite{Malti2021}.
Fig.~\ref{figMD6} shows the microstructural analysis of sintered Cu-Ni nanoparticles by the Common Neighbor Analysis (CNA) method. It can be found that when the sintering time increases, HCP structure gradually disappears. As a summary in Fig.~\ref{figMD6}(d), the amount of HCP and amorphous phase is remarkably decreased when the sintering duration is 100 ps~\cite{Malti2021}.

\begin{figure}[!b]
\centering
\includegraphics[width=8.4cm]{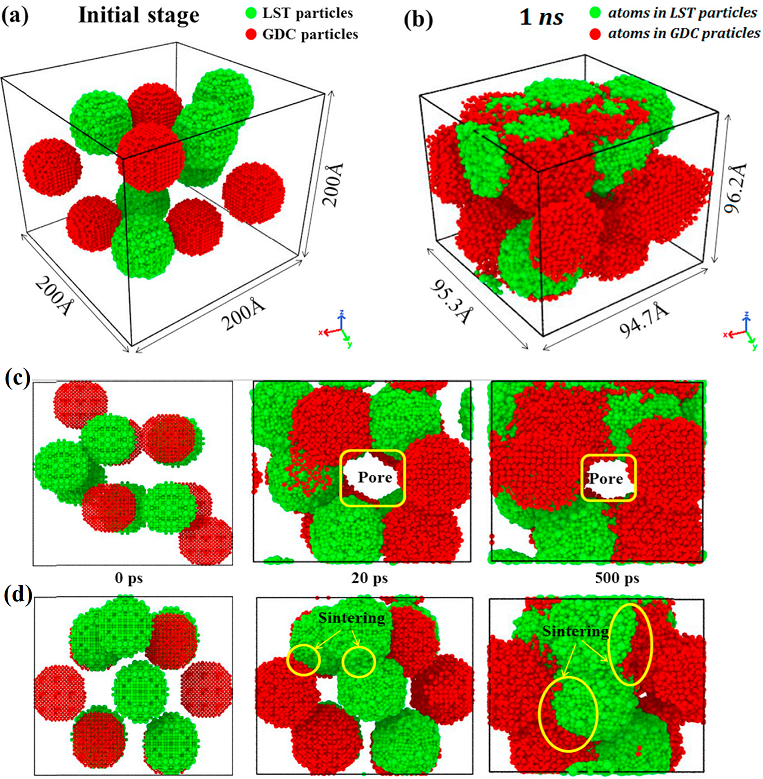}
\caption{(a) Initial and (b) final state of sintered LST-GDC nanoparticlesl. Cross-sectional view of the structures at different sintering time along the (c) $y$ and (d) $x$ direction of the simulation box. Reproduced with permission~\cite{Yang2021}. Copyright 2021, Elsevier.}
\label{figMD5}
\end{figure}

Recently, MD simulations are utilized to study the sintering process of composites such as the electrode of reversible solid oxide fuel cell (rSOFC) and graphene-metal composites.
In the rSOFC field, LST-GDC (LST: SrLaTiO; GDC: Gd-doped ceria), Ni-YSZ (YSZ: yttria-stabilized zirconia), and NiO-YSZ composites are promising candidates as electrodes. MD simulations of sintered LST-GDC~\cite{Yang2021}, Ni-YSZ \cite{Nakao2013Molecular,Nakao2013MoleularYSZ,Xu2013Molecular}, and NiO-YSZ~\cite{Fu2017} nanocomposites are carried out. For the LST-GDC nanoparticles sintering, it is revealed that a high sintering temperature favors the increase in length of the triple-phase boundary, but is not beneficial for the enhancement of effective surface area of catalyst particles.
Fig.~\ref{figMD5} presents the temporal evolution of LST-GDC microstructure. The initial and final multiple nanoparticles system that is sintered at 1673 K for 1 ns is shown in Fig.~\ref{figMD5}(a) and (b), respectively.
At 20 ps in Fig.~\ref{figMD5}(c) and (d), LST and GDC nanoparticles gather together due to attractions among each particles. Then a porous structure forms. At 500 ps, most small particles are deformed and further amalgamated into particles with larger size. The volume heat capacity, thermal conductivity, and thermal expansion coefficient calculated by MD simulations of the sintered products is revealed to agree well with the experimental results. Moreover, quiet a few LST-GDC contact boundaries are found to generate triple-phase boundaries. This is favorable for the chemical reactions where charge exchange predominates. These findings by MD simulations could guide the design of sintering processing parameters for LST-GDC composite electrodes in rSOFC technology.

\begin{figure}[!t]
\centering
\includegraphics[width=8.4cm]{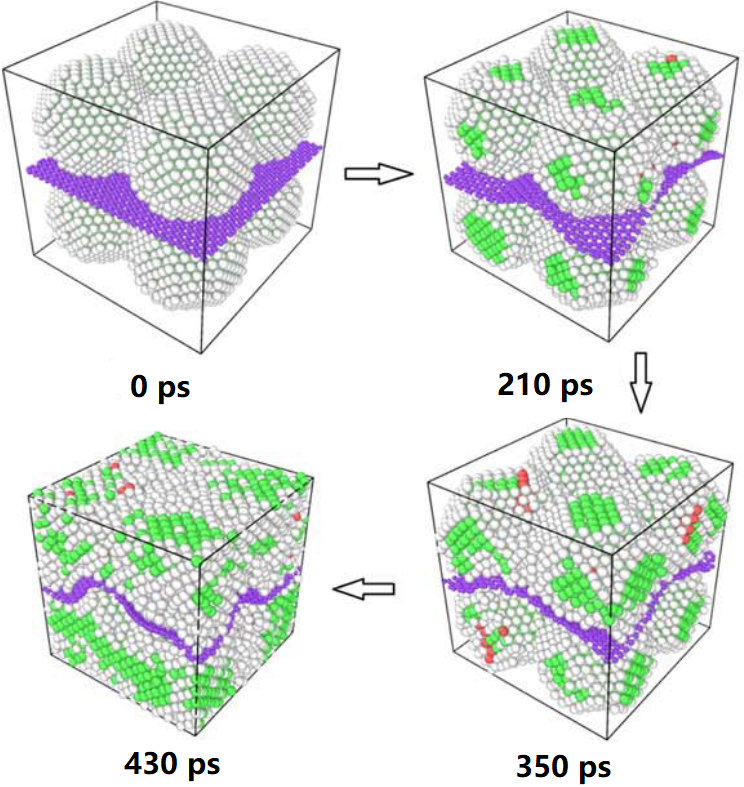}
\caption{Snapshots of composite system containing graphene nanoplatelet and Al nanoparticle sintered at 900 K. Reproduced with permission~\cite{He2019Molecular}. Copyright 2019, IOP.}
\label{figMD4}
\end{figure}

The sintering process of aluminum matrix powder reinforced by graphene nanoplatelet is recently explored by MD simulations~\cite{He2019Molecular,Zhu2022Atomistic}. As shown in Fig.~\ref{figMD4}, four spherical Al nanoparticles with an embedded graphene nanoplatelet constitute the cubic cell as the simulation box. The evolution of crystalline structures including FCC, HCP and others are recorded as a function of sintering duration in Fig.~\ref{figMD4}. The sintering mechanism is similar to that in the pure metallic nanoparticles. Additionally, sintering process makes graphene nanoplatelet wrinkled, leading to large contact area among Al nanoparticles and graphene nanoplatelet. It is further confirmed by MD simulations that the presence of graphene nanoplatelet can remarkably enhance the composites' mechanical properties, owing to the graphene-Al stress transfer and the reinforcement by dislocation~\cite{He2019Molecular}.

\subsection{Microstructure-scale method}
In contrast to the atomic-scale MD method which tracks the atomic trajectories to decipher the atomistic sintering mechanism, microstructure-scale method aims at the microstructure evolution in terms of grain size, grain shape, particle arrangement, impurities, lattice distortions, porosity, etc., which determine the properties or performance of the final sintered products. In general, microstructure-scale method depends on the constructed free energy, thermodynamic driving force, or particle-particle interactions.
In the following, discrete element method, Monte--Carlo method, and phase-field model will be overviewed.

\subsubsection{Discrete element method}
Discrete element method (DEM) is a numerical technique that calculates the interaction of a huge amount of particles and simulates the behavior of both continuous and discontinuous material systems \cite{Mahmood2016AReview,Harthong2009Modeling}. The basic theory of DEM is the Newton's second law of motion and rigid body dynamics, with the theoretical description of Newtonian interactions among particles. Constitutive behaviors have to be assigned to these particles, such as heat transfer models, contact models, collision models, inter-particle bond formation and break models, models for response to external fields, lase-matter interactions, etc~\cite{Onate2011Particle}. DEM is capable of calculating rotations and displacements of discrete bodies with different shapes. Within DEM, particles are numerically simulated by solving the governing equations via specific time-stepping algorithms~\cite{Kloss2011LIGGGHTS}. 

\begin{figure}[!bh]
\centering
\includegraphics[width=8.4cm]{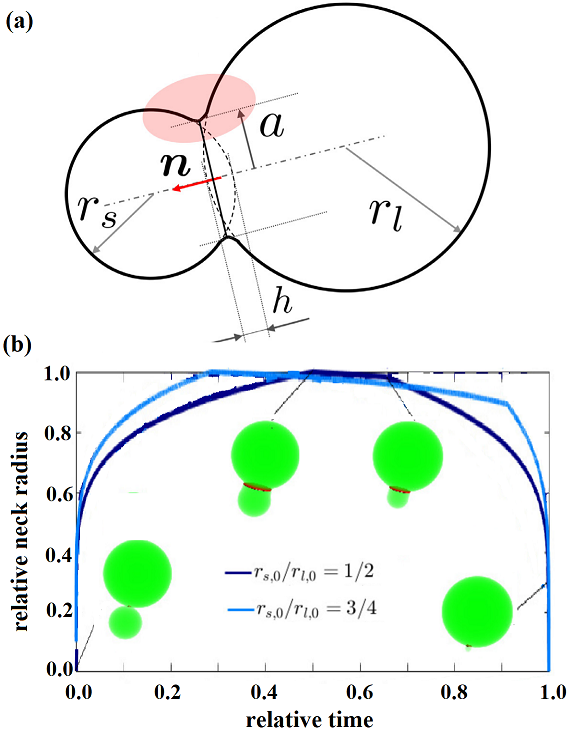}
\caption{(a) DEM geometrical parameters of a pair of particles. (b) DEM simulation results on the relative neck radius for the cases of two different initial size ratio $r_\text{s,0}/r_\text{l,0}$. Reproduced with permission~\cite{Paredes-Goyes2021}. Copyright 2021, Elsevier.}
\label{figDEM1}
\end{figure}

A general DEM implementation should consider the collision detection~\cite{Ericson2005Realtime,Ogarko2012Afast}, inter-particle contact laws~\cite{Kruggel-Emden2007Review}, boundary-condition contact laws~\cite{Martin2003Study}, bond formation, deformation, and breakage~\cite{Jiang2007Asimple}, time step integration~\cite{Kloss2011LIGGGHTS}, etc.
Unlike the classical DEM approach, during the sintering process the particle radius can evolve owing to the matter diffusion driven by the curvature gradient. When sintering is activated, mass transport due to the surface and grain boundary diffusion has to be considered. In the sintering case, Parhami and McMeeking~\cite{Parhami1998Anetwork} have proposed a particle contact force model that is originated from the calculations of Bouvard and McMeeking~\cite{Bouvard2005Deformation} for the particle pairs with identical size. Then Pan and his collegues~\cite{Pan1998Amodel} extended Bouvard and McMeeking's model to pairs of particles of different sizes. According to the expressions in these models, for a pair of particles with radii of $r_{s}$ and $r_{l}$, the normal force $N$ between two particles can be given as~\cite{Paredes-Goyes2021}
\begin{equation}
N = - \frac{\pi a^4}{(1+r_s/r_l)\beta \Delta_\text{GB}} \frac{\text{d}h}{\text{d}t} + \frac{\alpha}{\beta} \pi r_{l} \gamma_\text{s}
\end{equation}
in which $\gamma_\text{s}$ is the surface energy and the term related to diffusion is
\begin{equation}
\Delta_\text{GB} = \frac{\Omega}{k_\text{B}T} D_\text{GB} \delta_\text{GB}
\end{equation}
in which $D_\text{GB}=D_\text{0GB}\text{exp}(-Q_\text{GB}/RT)$ is the grain-boundary diffusion coefficient with an activation energy of $Q_\text{GB}$ at temperature $T$, $k_\text{B}$ the Boltzmann constant, $\delta_\text{GB}$ the thickness of grain boundary, and $\Omega$ the atomic volume. $\alpha$ and $\beta$ parameters are related to the ratio of the grain-boundary diffusion to surface diffusion. Other geometrical parameters are illustrated in Fig.~\ref{figDEM1}(a). If the grain growth mechanisms are activated during sintering, the volume evolution of a system containing a large particle $l$ that contacts a small particle $s$ is governed by
\begin{equation}
\frac{\text{d}V_{l,s}}{\text{d}t} = 4 \pi r_{l}^2 \frac{\text{d}r_{l}}{\text{d}t} = \sum_i J_i A_i \Omega .
\end{equation}
The volume change for a given contact is attributed to various mass transport mechanisms. The contribution is denoted by a flux cross-section area $A_i$ and an atomic flux density $J_i$ with $i$ representing the different mechanism, i.e., grain-boundary migration and surface diffusion. In this way, DEM model realizes a coupling between grain growth and sintering kinetics. Fig.~\ref{figDEM1}(b) shows the temporal evolution of neck radius during the sintering of a large and a small alumina particle. It can be found that DEM captures different sintering mechanisms at different stages. At the initial shrinkage stage, there is no grain growth and the neck is extremely small. At the intermediate stage, grains grow by surface diffusion, there is no apparent shrinkage, and the neck radius is not less than the equilibrium value. At the final stage, fast grain growth occurs by GB migration and there is apparent shrinkage when the neck radius is not less than the smallest particle radius~\cite{Paredes-Goyes2021}. Similar DEM simulations on the sintering behaviors of copper particles with the consideration of coarsening~\cite{Martin2006Discrete} are also carried out, providing a solid basis for the realistic simulation of coupling between coarsening phenomena and sintering by DEM.

In addition, DEM can not only investigate the microstructure evolution and anisotropic sintering of alumina particles~\cite{Rasp2012Shape}, but also be applied to the determination of density dependent anisotropic constitutive parameters of sintered alumina~\cite{Wonisch2007Stress}. 
In order to simulate long-time sintering by DEM, a relaxation time scale is introduced as a 'trick' to enable a long-time sintering simulation when the simulation time still remains small~\cite{Luding2005Adiscrete}.
Furthermore, for removing the limits from the usage of an explicit scheme where a very small time step has to be adopted during DEM sintering simulation, a non-smooth method, i.e., contact dynamics combined with a sintering contact law (in contrast to smooth dynamics), is proposed~\cite{Martin2014Simulation}. DEM simulation results of sintering indicate that contact dynamics is superior to smooth dynamics in terms of representing the rearrangement of particles~\cite{Martin2014Simulation}.

\begin{figure}[!b]
\centering
\includegraphics[width=8.4cm]{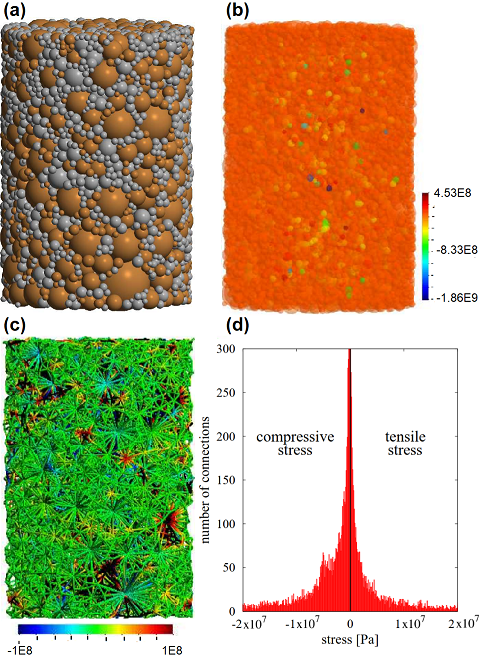}
\caption{(a) DEM model of two-phase 80\%NiAl-20\%Al$_2$O$_3$ powder. (b) Total hydrostatic stresses particles at the final-stage sintering. (c) Spatial distribution and (d) histogram of viscoelastic microscopic stresses at the final-stage sintering. Reproduced with permission~\cite{Nosewicz2020Discrete}. Copyright 2020, MDPI.}
\label{figDEM2}
\end{figure}

Simulations on the sintering behavior and controlled microstructures, with a focus on composite electrode, metallic matrix with ceramic inclusions or metal-ceramic composite, initial green microstructure, and pre-cracked plate, indicate that DEM possesses a huge potential in understanding the correlation between defect formation, residual stress and microstructure evolution in the sintering of diverse material systems~\cite{Martin2016Sintered, Beloglazov2020Discrete,Nosewicz2020Discrete,Besler2016Discrete}.
For instance, the viscoelastic DEM~\cite{Nosewicz2013Viscoelastic} model is employed to investigate the pressure-assisted sintering of intermetallic-ceramic composite NiAl-Al$_2$O$_3$ and the post-sintering residual stress field~\cite{Nosewicz2020Discrete}. Fig. \ref{figDEM2}(a) shows the DEM model geometry of a 80\%NiAl-20\%Al$_2$O$_3$ powder. The spatial distribution of particle total microscopic stresses and viscoelastic microscopic stresses at the final stage of sintering can be seen in Fig.~\ref{figDEM2}(b) and (c), respectively. The viscoelastic resistance of the material is balanced by the attractive contact interaction that is resulted from the sintering driving stress and external stress. The positive and negative values of viscoelastic microscopic stress in Fig.~\ref{figDEM2}(c) and (d)~\cite{Nosewicz2020Discrete} indicate the tensile and compressive interaction, respectively.

\begin{figure}[!th]
\centering
\includegraphics[width=8.4cm]{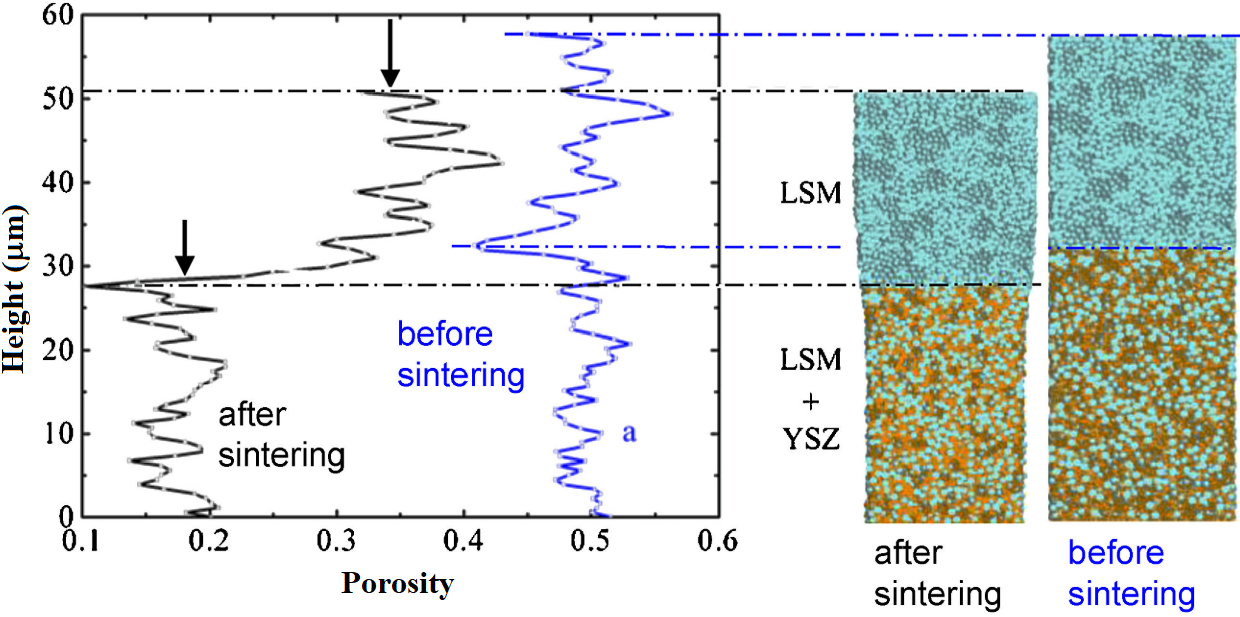}
\caption{DEM results on porosity distribution and microstructure of a bilayered YSZ-LSM composite electrode before sintering and after sintering. Reproduced with permission~\cite{Martin2016Sintered}. Copyright 2016, The Ceramic Society of Japan.}
\label{figDEM3}
\end{figure}

One of the recently attractive topics is the application of DEM to the sintering simulation of composite that is essential for energy storage. For example, Fig.~\ref{figDEM3} shows the DEM simulation results on the sintering of LSM-YSZ bilayer composite electrode (LSM: La$_{1-x}$Sr$_x$MnO$_3$) by introducing the pore formers. LSM-YSZ composites with tunable and controllable porosity are excellent candidates for SOFC cathodes. The top layer contains pore formers and LSM, while the bottom layer is LSM-YSZ composite. The initial green porosity of both layers is around 0.5. Owing to the smaller size of YSZ particles (0.5 $\mu$m), the eventual porosity of the bottom layer is remarkably decreased, whereas the top layer experiences relatively small porosity decrement. The interface distortion between bottom and top layers is also an indicator of higher densification in the bottom YSZ-LSM layer~\cite{Martin2016Sintered}. Specifically, DEM simulations of sintering a 40:60 vol ratio mixture of LSM and YSZ are performed to investigate the anisotropic sintering behavior. It is confirmed that sintering anisotropy occurs at the macroscopic scale, as well as within the macropores and walls. The aligned macropores could induce an anisotropic particle contact network and thus an anisotropic shrinkage during sintering~\cite{Lichtner2018Anisotropic}. As for the co-sintering of multilayer ceramic capacitors (e.g., Ni/BaTiO$_3$), a parametric study based on DEM simulations reveals the influence of green density, heating rate, and thickness, nickel particle rearrangement, and non-sintering inclusions on microstructure features of the final sintered electrode. The associated results are beneficial for the sintering design of multilayer capacitors~\cite{Yan2014Microstructure}.

\begin{figure}[!th]
\centering
\includegraphics[width=8.4cm]{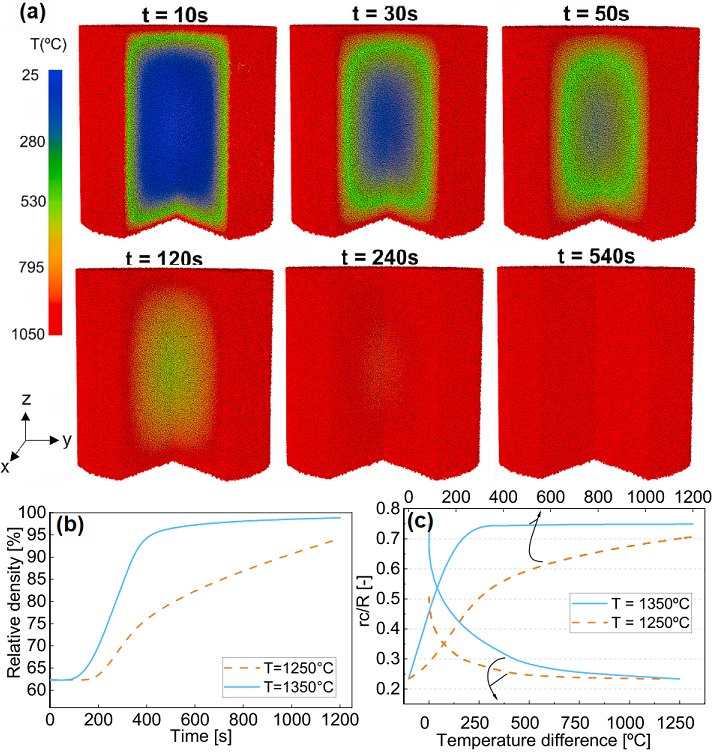}
\caption{(a) DEM results on profile of temperature evolution over the soaking time at 1050 $^\circ$C. Evolution of (b) relative density and (d) normalized average neck radius (r$_\text{c}$/R) over the temperature difference (bottom $x$-axis) and sintering time (upper $x$-axis) at 1250 and 1350 $^\circ$C. Reproduced with permission~\cite{Teixeira2021High}. Copyright 2021, Elsevier.}
\label{figDEM4}
\end{figure}

For the high heating rate sintering such as rapid firing, DEM model for coupling thermomechanics with sintering and non-isothermal solid-state sintering is proposed~\cite{Teixeira2021High}. Unlike the case of traditional sintering where the heating or cooling is extremely slow, the evolution of thermal and densification gradients is critical for the high heating rate sintering. The temperature gradient plays an important role in microstructure evolution.
As an example, the rapid sintering during the fast firing of Al$_2$O$_3$ particles is simulated by DEM. Fig.~\ref{figDEM4}(a) presents the temperature evolution history from 300 to 1050 $^\circ$C within around 200 s. The outer particles are immediately heated up to 1050 $^\circ$C in less than 10 s, while the inner particles remain at room temperature. Hence, there exists a very large temperature gradient. As shown the relative density evolution in Fig.~\ref{figDEM4}(b), even sintered at a lower temperature increase rate for the case of 1250 $^\circ$C,  the powders' densification is gradually increased and a maximum relative density of 94\% is obtained at 1200 s. The normalized average neck radius r$_\text{c}$/R as a function of temperature difference and the sintering time is presented in Fig.~\ref{figDEM4}(c). For the fast sintering at 1350 $^\circ$C, when the temperature difference is changed from 1250 to 0 $^\circ$C, a three-fold increase in r$_\text{c}$/R can be achieved. As the sintering temperature increases, the densification rate and the speed at which the temperature gradient passes through the sample are enhanced~\cite{Teixeira2021High}.

\begin{figure}[!b]
\centering
\includegraphics[width=8.4cm]{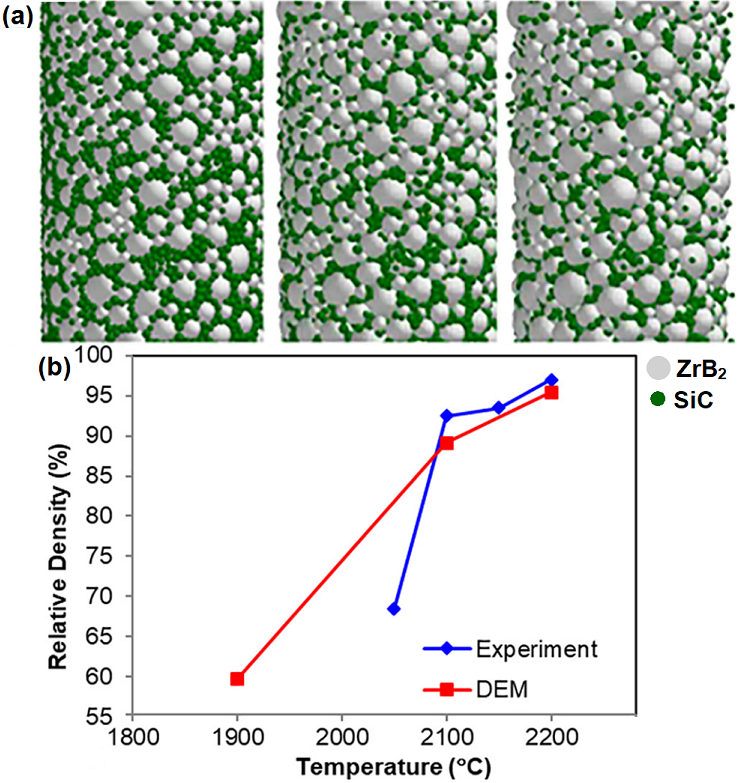}
\caption{(a) DEM results on the microstructure during the pressureless sintering process of ZrB$_2$-15 wt\%SiC composite at 2200 $^\circ$C. (b) Densification results of ZrB$_2$-15 wt\%SiC composite at different sintering temperatures. Reproduced with permission \cite{Iacobellis2019Discrete}. Copyright 2019, Elsevier. Experimental results in (b) reproduced with permission~\cite{Mashhadi2015Pressureless}. Copyright 2015, Elsevier.}
\label{figDEM5}
\end{figure}

The application of DEM to the simulation of sintering of ultra-high temperature ceramics (UHTCs) has recently received wide interests, due to the potential use of UHTCs in extreme environment conditions such as space travel, aerospace, and nuclear industry. In particular, DEM simulation of sintering of a typical UHTC consisting of ZrB$_2$ and SiC is reported~\cite{Iacobellis2019Discrete}. The DEM approach combined with a viscoelastic/sintering model is employed to study the sintering of ZrB$_2$-SiC composites. The interacting forces between particles, which is originated from the surface diffusion and grain boundary diffusion, is considered. The viscoelastic effect is included to ensure sufficient mass transport among interacting particles. Fig.~\ref{figDEM5}(a) shows the DEM simulation results about the sintered microstructure at the beginning, intermediate, and end of the sintering process of ZrB$_2$-15 wt\%SiC composite at 2200 $^\circ$C. Since there exists an initial interaction/compaction with the wall when the Hertz--Mindlin model is used, it is clear that some particles close to the wall are detached and dangle in the air. Owing to the lattice diffusion and grain boundary migration, the main body shrinks. Fig.~\ref{figDEM5}(b) presents the final relative density at different sintering temperatures. It can be seen that higher sintering temperature favors higher densification. When the sintering temperature is increased above 1900 $^\circ$C, the densification is significantly enhanced. This is foreseeable, since UHTCs retain the strengths and structure of solid phase up to 2000 $^\circ$C~\cite{Iacobellis2019Discrete}. The DEM simulation results are found to agree well with the experimental measurement~\cite{Mashhadi2015Pressureless}.

\begin{figure}[!t] 
\centering
\includegraphics[width=8.4cm]{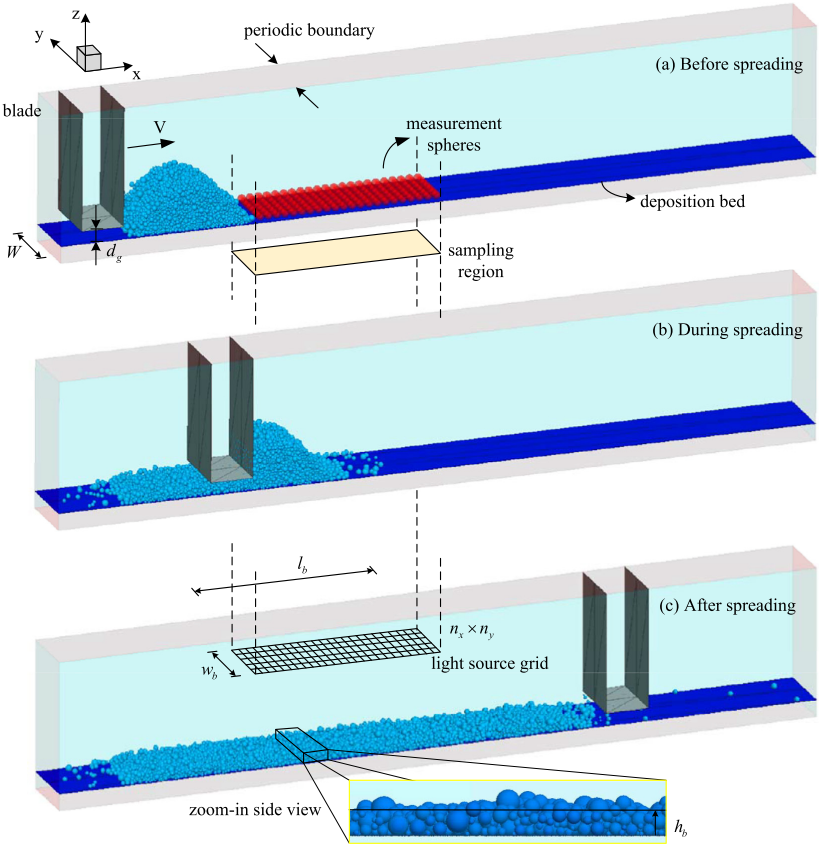}
\caption{DEM setup of spreading powder for additive manufacturing. Reproduced with permission~\cite{Ma2020Numerical}. Copyright 2020, Elsevier.}
\label{figDEM80}
\end{figure}

Another recently flourish topic is the application of DEM to the sintering based additive manufacturing.
As an essential process of powder-based additive manufacturing, powder-bed spreading has been widely modelled and simulated by DEM ~\cite{Shaheen2021,Haeri2017Optimisation,Nan2019,Chen2019Powder,
Ma2020Numerical,Wang2020Adhesion,Zhao2017Characterization,Markl2018Powder}. The influence of surface roughness, complex particle shape, particle cohesion, and spreading tool type on the powder-layer quality is examined by DEM~\cite{Shaheen2021,Wang2020Adhesion}. A compromise between material parameters and processing parameters is found. If the the spreading speed is increased, the layer quality of strongly cohesive powders and non-cohesive (or weakly cohesive) powders would be increased and decreased, respectively.
DEM simulations are also applied to optimize the blade type of spreaders for the generation of powder bed~\cite{Haeri2017Optimisation}. It is shown that if the geometry of a blade spreader is optimized, the quality of a powder bed with higher solid volume fraction and lower surface roughness would be notably improved. Based on DEM results, a new spreading device with a super-elliptic edge profile which has three parameters to control height, width, and overall shape is proposed.
The fundamental mechanisms of the powder-layer packing during powder-spreading process are also explored by DEM~\cite{Chen2019Powder,Ma2020Numerical}. It is found that there exist three kinds of deposition mechanisms: wall effect, cohesion effect, and percolation effect. These mechanisms compete with each other and finally determine the powder-layer packing density. At the bottom of powder pile that is scraped by the rake, a stress-dip is found to make the powder particles uniformly deposited~\cite{Chen2019Powder}.
In particular, the influence of fine fraction on the flow ability of powders in additive manufacturing is examined by DEM simulations, with a focus on the microscale mechanisms of powder-bed flow that involves van der Waals force~\cite{Ma2020Numerical}. Firstly, particles are randomly generated in a cube to prepare the initial spreading configuration. Then, the particles are deposited under gravity to reach an equilibrium state, as shown in Fig.~\ref{figDEM80}. It is found that the flow ability of coarse particles is hardly influenced by the microscale force. However, the microscale force controls the behavior of fine fractions of micrometer-scale particles~\cite{Ma2020Numerical}.


\begin{figure}[!t]
\centering
\includegraphics[width=8.4cm]{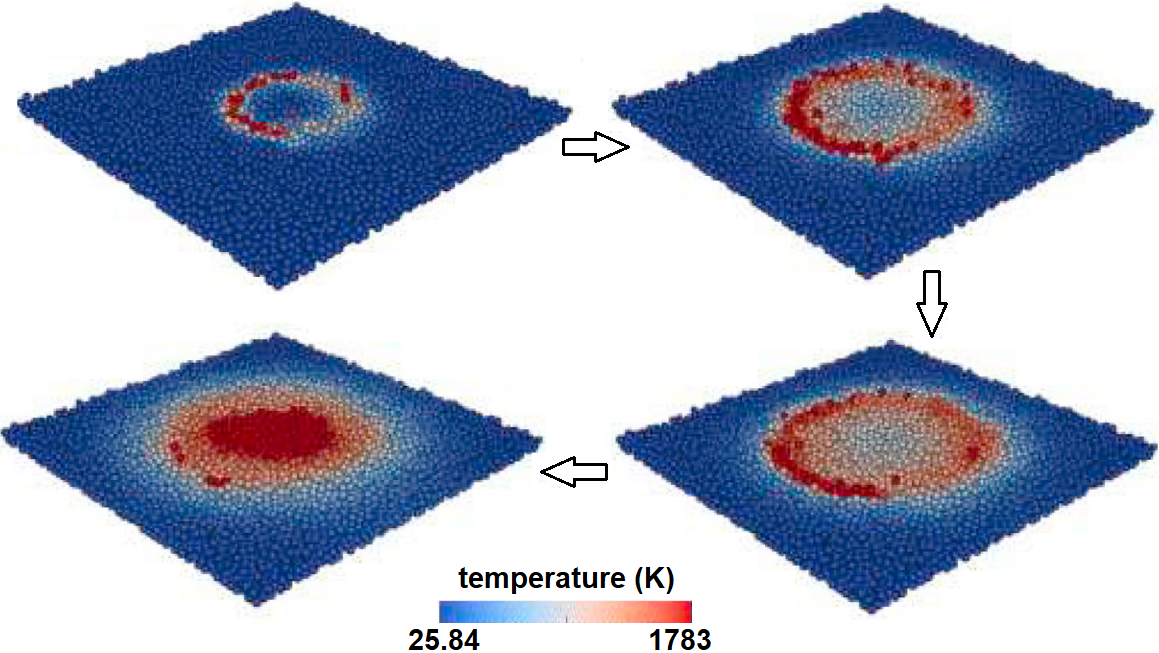}
\caption{DEM simulation results on the laser-sintering printing of inner and outer regions of a cylinder. Reproduced with permission~\cite{Xin2018Discrete}. Copyright 2018, Elsevier.}
\label{figDEM6}
\end{figure}

\begin{figure}[!t]
\centering
\includegraphics[width=8.4cm]{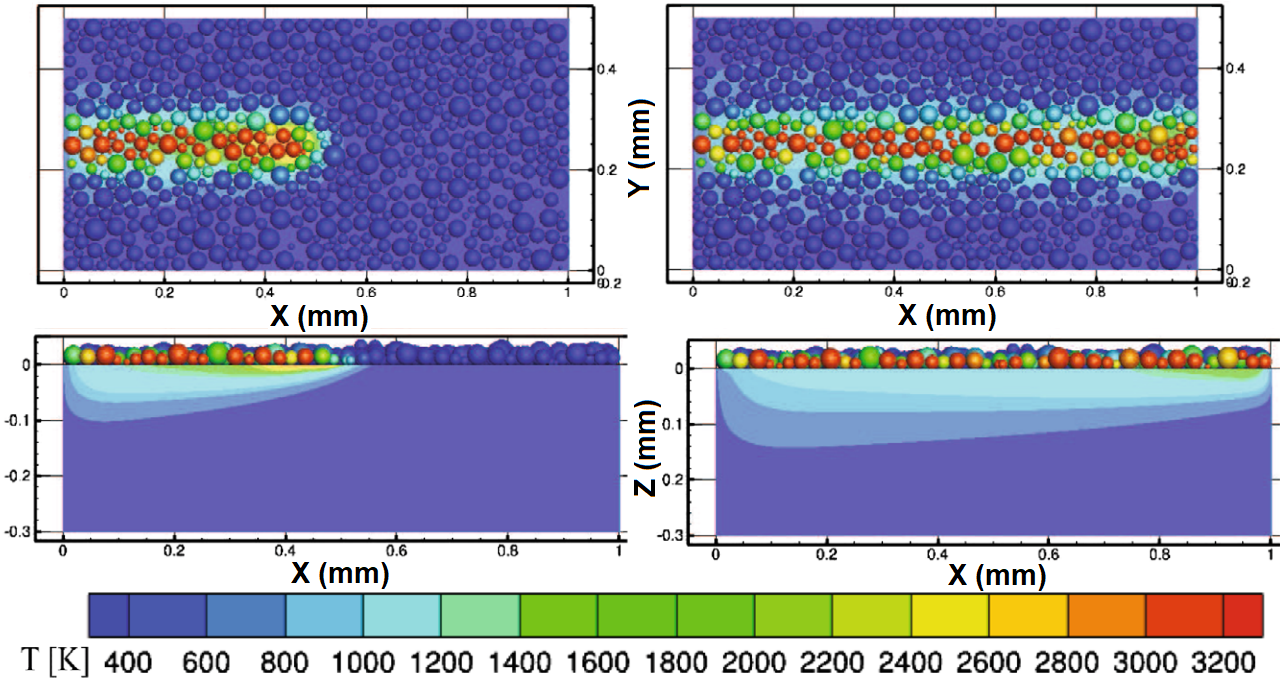}
\caption{DEM simulation results on the temperature distribution in 316L stainless steel particles and the underlying substrate after a laser scan is finished. Reproduced with permission~\cite{Ganeriwala2016}. Copyright 2016, Springer.}
\label{figDEM8}
\end{figure}

\begin{figure}[!t]
\centering
\includegraphics[width=8.4cm]{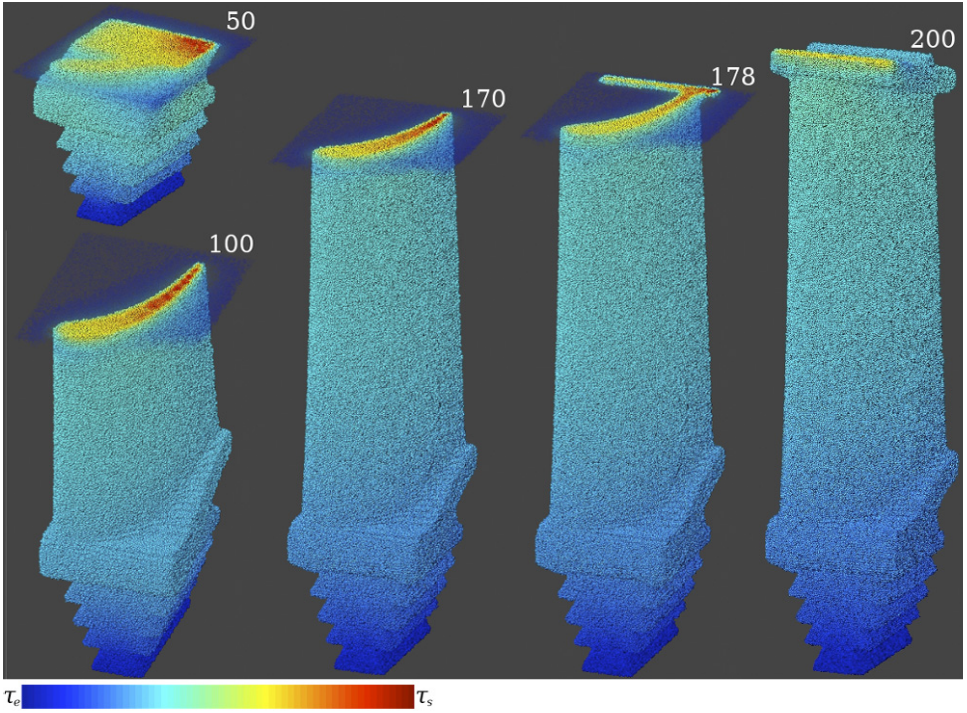}
\caption{DEM simulation results on the jet turbine blade's geometry and temperature distribution at the 50$^\text{th}$, 100$^\text{th}$, 170$^\text{th}$, and 178$^\text{th}$ layers. Reproduced with permission~\cite{Steuben2016}. Copyright 2016, Elsevier.}
\label{figDEM7}
\end{figure}

The sintering-based additive manufacturing process and the evolution of involved physical quantities can also be simulated by DEM. A novel DEM is developed to simulate the whole direct metal laser sintering (DMLS) process, including the powder deposition, recoating, laser heating, and holding stages~\cite{Lee2017Discrete}. Using DEM, the influence of laser scanning speed, laser power, and hatch spacing on the powder-bed temperature distributions can be predicted during the DMLS process. Furthermore, DEM is improved to consider phase transformation, heat conduction, and inter-particle sintering (bond neck growth), in order to more accurately simulate the sintering-based additive manufacturing process of powder bed~\cite{Xin2018Discrete}. Fig.~\ref{figDEM6} presents several snapshots for the first-layer printing process, showing the temperature distribution during the laser-sintering printing of inner and outer regions of a cylinder~\cite{Xin2018Discrete}. Specifically, a methodology that couples DEM and finite-difference method for simulating SLS is proposed~\cite{Ganeriwala2016}. Therein, the powder particles are modelled as  thermally and mechanically interacting discrete spheres and the substrate is modelled via the finite-difference method. Fig.~\ref{figDEM8} presents the temperature evolution in particles and substrate during the laser scanning process.
For simulating the additive manufacturing processes where the radiative phenomena are of critical importance, a DEM is developed to account for the heat diffusion and radiative transfer by using a modified Monte Carlo-ray tracing method and the Mie scattering theory~\cite{Xin2017Numerical}. DEM incorporating thermal physics has been applied to simulate the layer-by-layer additive manufacturing processes of metal powders~\cite{Steuben2016,Steuben2016OnMultiphysics}. As shown in Fig.~\ref{figDEM7}, the SLS additive manufacturing process of a small jet turbine blade is simulated by DEM~\cite{Steuben2016}.

\begin{figure}[!t]
\centering
\includegraphics[width=8.4cm]{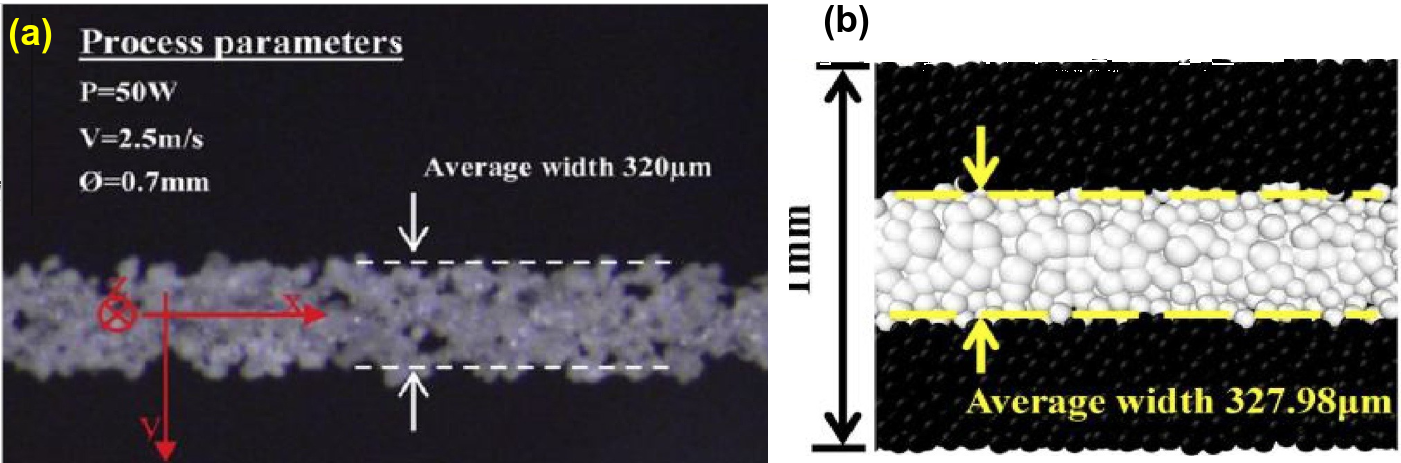}
\caption{
(a) Experimentally and (b) numerically (by DEM combining with a modified Monte--Carlo ray-tracing method) obtained microstructure in a polymer-SLS line. Reproduced with permission~\cite{Xin2017Multiphysicalmodel}. Copyright 2017, Elsevier.
}
\label{figDE-polymer}
\end{figure}

DEM combining with a modified Monte--Carlo ray-tracing method has been developed to simulate the multiphysical behavior of polymer particles during SLS process~~\cite{Xin2017Multiphysicalmodel}. The comprehensive model couples the underlying physics and the corresponding numerics, and could consider heat conduction, radiative heat transfer, sintering and granular dynamics. The simulated sintered lines of polymer powder after single-track scanning is shown in Fig.~\ref{figDE-polymer}(b), whose average width (320 $\mu$m) agrees well with that from experiments (328 $\mu$m) in Fig.~\ref{figDE-polymer}(a).

\subsubsection{Monte--Carlo method}
Monte--Carlo method is another popular method for the simulation of microstructure evolution, grain growth, and re-crystallization of polycrystalline materials.
Anderson \textit{et al.}~\cite{Anderson1984Computer} in 1984 for the first time proposed a novel Monte--Carlo procedure to investigate the grain growth kinetics in two dimensions. 
This procedure is based on the standard Q-state Potts model~\cite{Anderson1984Computer,Tikare2003}. In detail, the microstructure in the model is mapped onto a discrete lattice. A number $q_i$ (1$\geq q_i \leq$Q) is assigned to each lattice site $i$, representing the local crystallographic orientation. The initial orientations are set to randomly distribute. The total energy for the model is expressed as
\begin{equation}
E_\text{MC} = \frac{1}{2} \sum_{i=1}^N \sum_{j=1}^n J_{ij} \left[ 1 - \delta(q_i, q_j) \right]
\end{equation}
in which $N$ is the total number of sites in the model system, $n$ is the number of nearest neighbor sites of site $i$, $J_{ij}$ is the interaction energy between $i$ and $j$ sites, and $\delta$ is the Kronecker delta with $\delta(q_i,q_j)=1$ for $q_i=q_j$ and $\delta(q_i,q_j)=0$ for $q_i \ne q_j$. In the grain growth algorithm for sintering, a grain site is exchanged with one of the neighboring grain sites chosen randomly. Using a standard Metropolis algorithm, the exchange is accepted with a probability $P$
\begin{equation}
P = \left\{ \begin{aligned}
& \text{exp}\left(- \frac{\Delta E_\text{MC}}{k_\text{B}T}\right), \quad  \Delta E_\text{MC} > 0 \\
& 1, \quad \quad \quad \quad \quad \quad \; \; \Delta E_\text{MC} \le 0
\end{aligned}
\right.
\label{eqP}
\end{equation}
in which $\Delta E_\text{MC}$ is the energy change by the exchange trial, $k_\text{B}$ is Boltzmann constant, $T$ is the simulation temperature. In the algorithm for simulating pore migration, a pore site is interchanged with a neighboring grain site, and the acceptance probability can be similarly formulated according to Eq.~(\ref{eqP}).

\begin{figure}[!t]
\centering
\includegraphics[width=5cm]{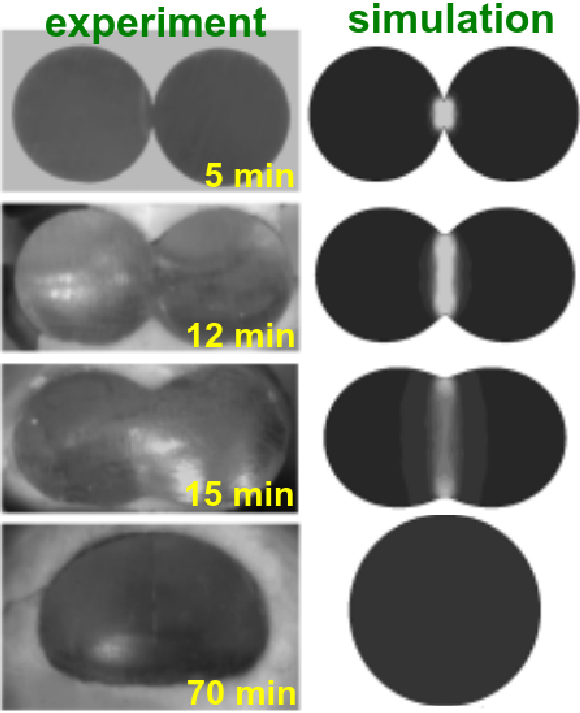}
\caption{Morphological evolution in two-glass cylinders sintered at 950 $^\circ$C: comparison between experimental observation and Monte--Carlo simulation. Reproduced with permission~\cite{Bordered2005Monte}. Copyright 2005, Wiley.}
\label{figMCexp}
\end{figure}

\begin{figure}[!t]
\centering
\includegraphics[width=8.4cm]{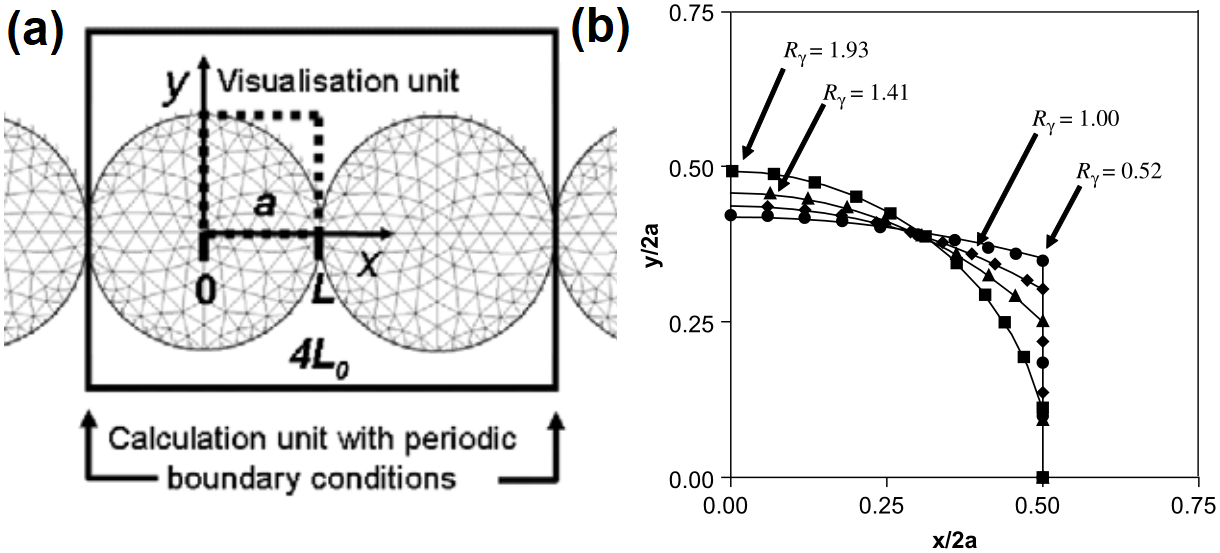}
\caption{(a) Initial configuration for simulating the sintering of an infinite row of cylinders with periodic boundary condition. (b) The stable morphologies after sintering from Monte--Carlo simulations (symbols) and from analytical resolution (solid line) for the constrained sintering ($L$ in (a) remains constant) in which $R_\gamma$ is the ratio of grain boundary energy to surface energy. Reproduced with permission~\cite{Bordere2006Improvement}. Copyright 2016, Elsevier.}
\label{figMC1}
\end{figure}

As an ideal case, Monte--Carlo simulations for sintering two or three particles are widely carried out to validate the simulation methodology and understand the sintering mechanism~\cite{Tikare2003,Bordere2008,Bordere2006}. For instance, in the sintering of a two-dimensional system containing three particles, Monte--Carlo model is shown to be capable of capturing the curvature-driven pore migration, grain growth and coarsening, as well as the densification by vacancy annihilation and vacancy diffusing to grain boundaries~\cite{Tikare2003}. The model is also verified by morphologic changes and densification kinetics. For the case of viscous sintering of two particles, an energetic potential based non-discrete Monte--Carlo (NDMC) methodology is proposed. The NDMC kinetics is demonstrated to agree well with the viscous sintering kinetics~\cite{Bordered2005Monte,Bordere2006,Bordere2008}.
Fig.~\ref{figMCexp} presents the experimental observation and NDMC simulation on the morphological evolution in two-glass cylinders sintered at 950 $^\circ$C at different times, indicating the consistency between experimental and simulations results~\cite{Bordered2005Monte}.
NDMC methodology is further employed to the free or constrained sintering of an infinite row of particles in two dimension, as shown in Fig.~\ref{figMC1}. The model geometry for simulating the sintering of an infinite row of cylinders with periodic boundary conditions along $x$ direction is presented in Fig.~\ref{figMC1}(a). The comparison between the final sintered morphologies  predicted by analytic solutions and Monte--Carlo simulations for different ratios ($R_\gamma$) of grain boundary energy to surface energy is given in Fig.~\ref{figMC1}(b) for the constrained sintering. It can be seen that the calculated interface morphologies match perfectly those from the analytical methods~\cite{Bordere2006Improvement}.

\begin{figure}[!t]
\centering
\includegraphics[width=8.4cm]{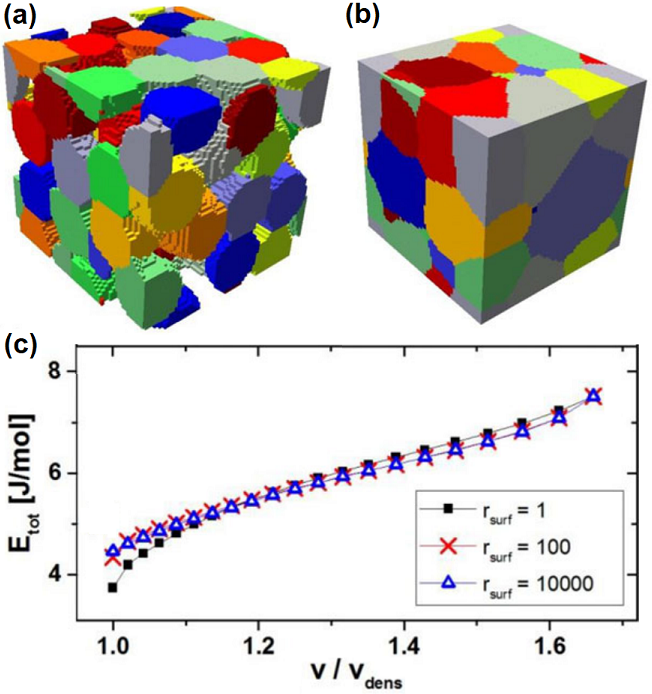}
\caption{Monte--Carlo simulations of sintering 50 particles at various fractional densities: (a) 70\% and (b) 100\%. (c) Total interface energy $E_\text{tot}$ \textit{vs} scaled specific volume ($v/v_\text{dens}$) for various frequencies of surface diffusion $r_\text{surf}$. Reproduced with permission~\cite{Raether2018Anovel}. Copyright 2018, Wiley.}
\label{figMC2}
\end{figure}

Monte--Carlo simulation methodology is also applicable to the sintering of many-particle systems or polycrystalline materials~\cite{RaoMadhavrao1989Monte,Akhtar1994Monte,
Zeng1998Potts,Morhac2000,Braginsky2005Numerical,Neizvestny2007,Kim2009,Raether2018Anovel}.
A Monte--Carlo technique is put forward for the initial-stage sintering of two-dimensional aggregates of copper particles which are randomly packed~\cite{RaoMadhavrao1989Monte}. Therein, the crack initiation that is attributed to the stresses generated by the sintering particles, as well as the localized stresses and stress evolution, can be considered. 
Similarly, Monte--Carlo simulation has been carried out to describe the gas phase coagulation and sintering of two-dimensional nano-clusters, in which the finite inter-particle binding energy, particle restructuring, and densification (sintering) are incorporated~\cite{Akhtar1994Monte}. 
In addition to the initial-stage sintering, for simulating the grain size distributions during the final-stage sintering, a Potts Monte--Carlo model is proposed to simultaneously obtain the pore migration, grain growth, and pore shrinkage~\cite{Zeng1998Potts}. The model could simulate a system with an initial porosity and varying ratios of grain boundary mobility to pore shrinkage rates.
For the sintering of a real polycrystalline material, a Monte--Carlo model describing monophase or two-phase structures is developed to deal with the oriented and anisotropic grain growth~\cite{Morhac2000}. Meanwhile, the model includes a considerable number of input parameters that make it possible to design lots of combinations of conditions under which the sintering process occurs.
For simulating microstructural evolution as well as macroscopic deformation during sintering of complex powder compacts, a model based on the kinetic Monte--Carlo approach is developed to capture vacancy diffusion, grain growth, pore annihilation at grain boundaries, and sintering stress~\cite{Braginsky2005Numerical}.
Different from these models, recently a novel Monte--Carlo model that considers local interface curvature in the acceptance of diffusion steps is developed, enabling the study of interface energy influence on sintering process~\cite{Raether2018Anovel}. Fig.~\ref{figMC2}(a) and (b) shows an example Monte--Carlo simulation result on the microstructure evolution during sintering at a dihedral angle of 120$^{\circ}$, with intermediate states at a fractional density of 70\% and 100\%. As shown in Fig.~\ref{figMC2}(c), for the final-phase densification where the scaled specific volume $v/v_\text{dens}$ approaches 1, the surface diffusion ($r_\text{surf}=1$) reduction would decrease the total interface energy . At high rate of surface diffusion (e.g. $r_\text{surf} = 10,000$), grain growth is suppressed and the particle diameter remains nearly constant up to the final density~\cite{Raether2018Anovel}.

\begin{figure}[!t]
\centering
\includegraphics[width=8.4cm]{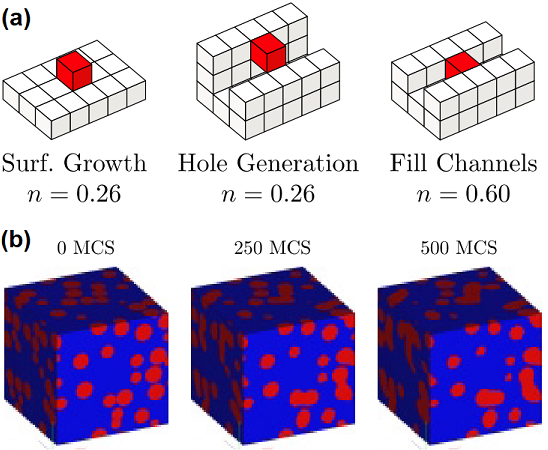}
\caption{A geometrical Monte--Carlo model for liquid-phase sintering: (a) Key configurations for generating the distribution function that is used to accept or reject voxel solidifications. (b) A complex microstructure evolution at isothermal conditions and different Monte--Carlo steps. Reproduced with permission~\cite{Luque2005}. Copyright 2005, IOP.}
\label{figMC3}
\end{figure}

For the special case of liquid-phase sintering, Monte--Carlo method is also applicable~\cite{Luque2005,Liu2006Therelation}. A geometrical Monte--Carlo model assuming constant temperature and homogeneous composition of liquid phase is proposed to simulate the microstructure evolution during liquid-phase sintering. The model works for the prediction of probabilities of solidification or melting by considering the local geometry via the closest surrounding neighbours and avoiding the use of thermodynamic probabilities~\cite{Luque2005}. 
The model's algorithm works on microstructures discretized by homogeneous cubic elements (voxels). The model only considers two kinds of neighbours: edge and face neighbours. A weight factor is defined for each kind of neighbour, with a normalized weight $n$ ranging from 0 to 1. As shown in Fig.~\ref{figMC3}(a), solidification of a voxel at a flat-surface top corresponds to the growth of flat surfaces with $n$ as 0.26. The generation of a hole also corresponds to $n=0.26$. But the filling of a channel gives $n$ as 0.60. Based on this strategy, the model is applied to the isothermal simulation of a complex 3D microstructure in Fig.~\ref{figMC3}(b). The initial microstructure is generated by growing 146 spherical seeds until they occupy 30\% of the volume. After several hundreds Monte--Carlo steps, microstructure smoothly evolves and particles tend to fuse and create large solid clusters~\cite{Luque2005}.
The effect of wetting angle on dihedral angle distribution and grain boundary penetration degree of a liquid phase during the liquid phase sintering is examined by Monte--Carlo simulations. It is found that in the case of high probability for wetting, the number of the grain boundary penetration of a liquid phase increases while the wetting angle decreases~\cite{Liu2006Therelation}.

\begin{figure}[!t]
\centering
\includegraphics[width=8.4cm]{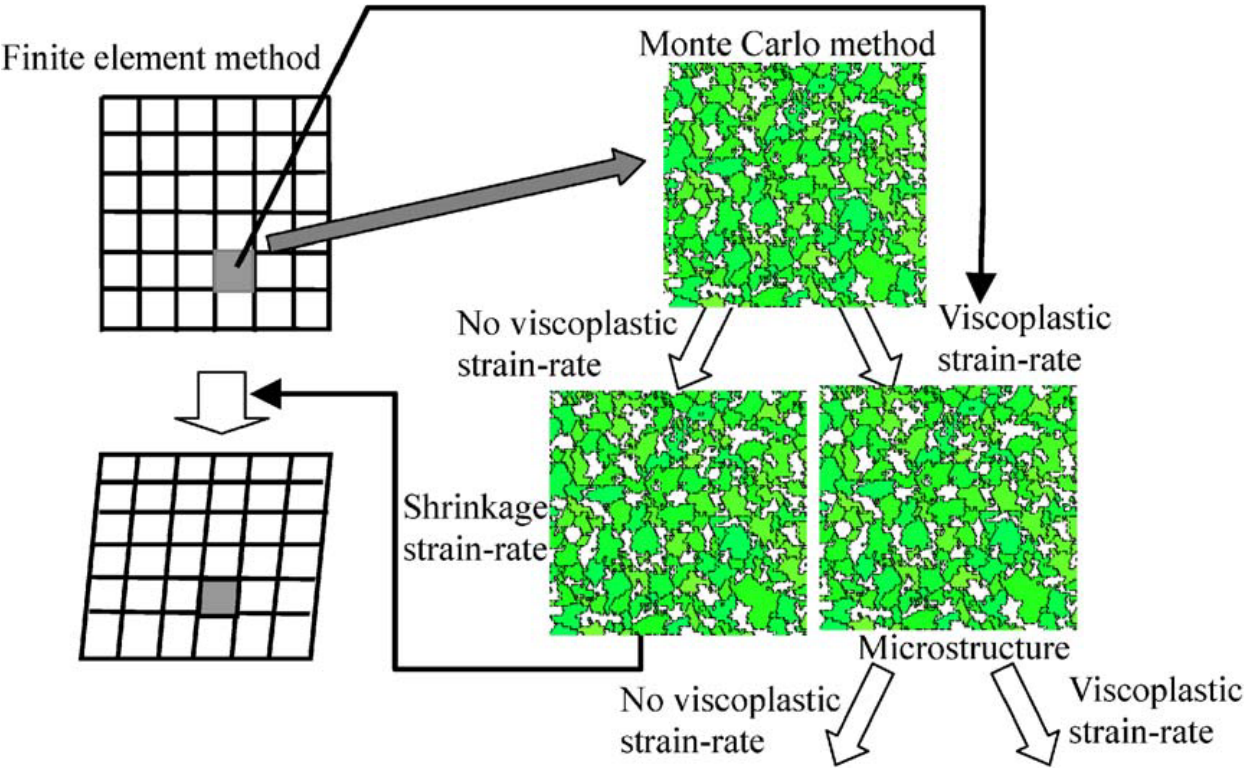}
\caption{Illustration on micro-macro simulations of sintering by combining finite element and Monte--Carlo methods. Reproduced with permission~\cite{Mori2004Micromacro}. Copyright 2004, Elsevier.}
\label{figMC4}
\end{figure}

\begin{figure}[!t]
\centering
\includegraphics[width=8.4cm]{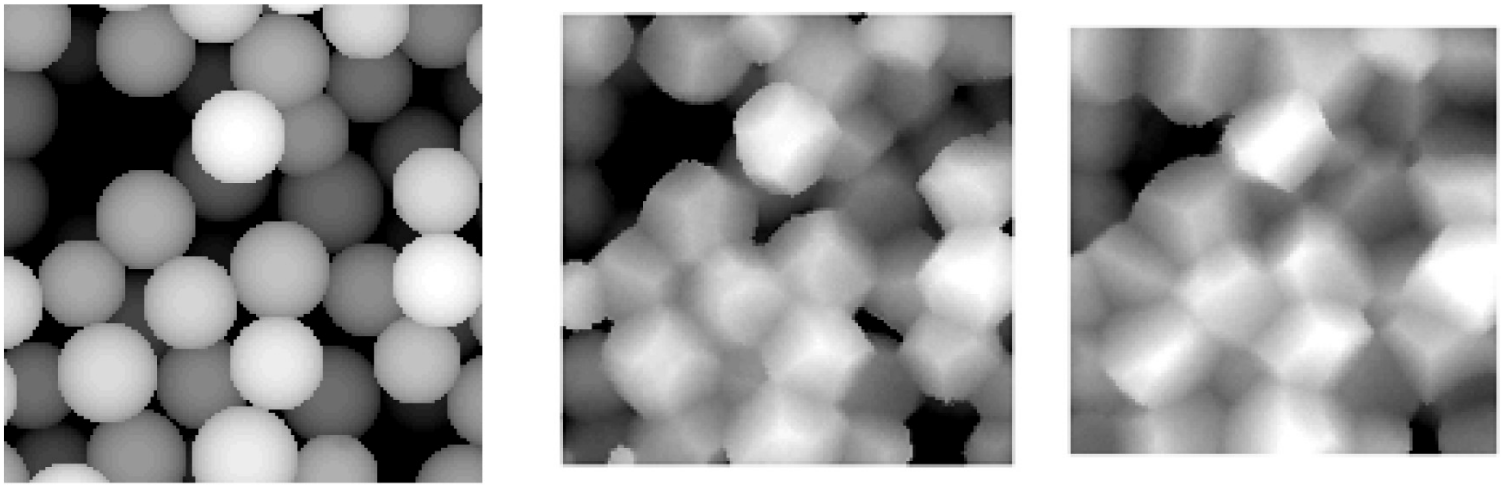}
\caption{Hybrid Cellular Automata--Monte Carlo simulations of the microstructure evolution at different sintering stages for spherical particles that are randomly packed. left: initial state; middle: 68 time steps; right: 300 time steps. Reproduced with permission~\cite{Wang2018Combining}. Copyright 2018, Elsevier.}
\label{figMC5}
\end{figure}

There are also attempts to combine Monte--Carlo method with other methods (e.g. finite element methods~\cite{Mori2004Micromacro}, Cellular Automata~\cite{Wang2018Combining}, etc.) for the simulation of sintering. For instance, a micro-macro method that combines finite element and Monte--Carlo methods is proposed to simulate the sintering process of ceramic powder compacts~\cite{Mori2004Micromacro}, as illustrated in Fig.~\ref{figMC4}. The compact body in the macroscopic scale is divided into finite elements, while the Monte--Carlo simulation is performed at the centre of each finite element. Firstly, Monte--Carlo simulations are performed to calculate the shrinkage strain rate without the consideration of macroscopic viscoplastic strain rate. Secondly, with the calculated shrinkage strain rate for each finite element as input, finite element simulations are carried out to attain the viscoplastic strain rate. Thirdly, Monte--Carlo simulations for the next time step is carried out with and without the inclusion of macroscopic viscoplastic strain rate. Then, the shrinkage strain rate and microstructure are attained without and with the macroscopic viscoplastic strain rate, respectively. The finite element and Monte--Carlo simulations are repeated in every time step~\cite{Mori2004Micromacro}. In addition, a hybrid Cellular Automata--Monte Carlo (CA--MC) approach is also proposed for the simulation of sintering process. The approach minimizes the energy dissipation rate and stored energy by using the variational principle to incorporate the influence of stress that is originated from the neighbouring particles/grains interactions. It could deal with grain boundary diffusion and collapse, and is demonstrated to be capable of simulating the sintering process of a randomly packed assembly of spherical particles, as shown in Fig.~\ref{figMC5}~\cite{Wang2018Combining}.

\begin{figure}[!t]
\centering
\includegraphics[width=8.4cm]{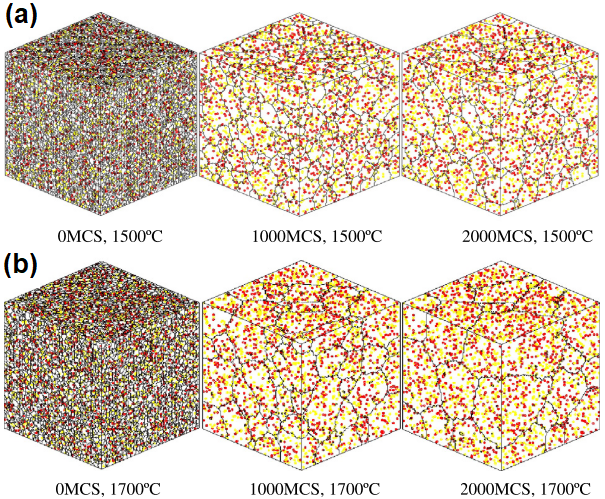}
\caption{Monte--Carlo simulations on microstructure of nano-composite ceramic materials at a sintering temperature of (a) 1500$^\circ$C and (b) 1700$^\circ$C. Reproduced with permission~\cite{Hao2011}. Copyright 2011, Elsevier.}
\label{figMC6}
\end{figure}

Other works are focused more on the application of Monte--Carlo method to the simulation of sintering of different materials such as Ni-YSZ composite~\cite{Hara2013}, nanoparticles~\cite{Guo2012Sintering,Qiu2008}, ceramics~\cite{Hao2011,Cheng2009,Zhao2008,
Suzuki1999Microstructural,Keum2002Computer,Matsubara1999Computer,Gu2017Simulation}, etc. Monte--Carlo method is utilized to predict the microstructure evolution of the Ni-YSZ composite system and reveal the effect of rigid YSZ phase on Ni coarsening kinetics~\cite{Hara2013}. Monte--Carlo simulations based on multi-state Potts model is applied to simulate the sintering of nano-particles by boundary migration, evaporation and condensation, in which the reduced temperature, the next-nearest neighbour weighting and the ratio of interface-surface energy ratio are taken as model variables~\cite{Qiu2008}. The sintering of ceramics such as Si$_3$N$_4$~\cite{Zhao2008} and Al$_2$O$_3$~\cite{Suzuki1999Microstructural,
Keum2002Computer,Gu2017Simulation} is widely simulated by Monte--Carlo method. In particular, the microstructure evolution of a three-phase nano-composite ceramic tool material during sintering process is simulated by a 3D Monte--Carlo model~\cite{Hao2011} at different sintering temperatures, as shown in Fig.~\ref{figMC6}. The model accounts for the grain boundary energy of each phase and interfacial energy between two phases as the driving forces for grain growth. It is found that the grain size is distributed more uniformly at 1700$^\circ$C than at 1500$^\circ$C.
In particular, Weiner et al.~\cite{Weiner2022ANew} proposed a new approach for sintering simulation of irregularly shaped powder particles. They combined Monte--Carlo approach and a statistical approach for describing particles’ morphology in a powder mixture.

\subsubsection{Phase-field model}
Phase-field model is an powerful methodology for the simulation and prediction of microstructure or morphology evolution. It adopts a set of conserved or non-conserved order parameters, which are continuous across the interface, to describe the microstructure~\cite{Chen2002,Chen2021From}. The spatial and temporal evolution of order parameters
represents the evolution of the microstructure. In general, phase-field theory is in the accordance with the variational theory. With the microstructure represented by the order parameters, the bulk energy/potential contributions are readily recaptured using the thermodynamics data, while energy/potential penalties are enforced at interface usually through the square of the order parameter gradient. In other words, the energy/potential of the system can be given as a functional of the order parameters. The evolution of microstructure can be formulated by the first variational derivative of the functional with respect to the order parameters. For the conserved order parameters, the Cahn--Hilliard equation is generally applied, while for non-conserved ones the Allen--Cahn equation is applicable, with the corresponding kinetic parameters involved. Phase-field models are able to readily capture the evolution of complex microstructure and arbitrary morphology without the explicit tracking of interface.
There are several merits of phase-field model~\cite{Qin2010}. Firstly, there is no need to develop special numerical techniques to explicitly track the interface position. Major parameters in the energy functional can be derived from physical quantities, for instance, interface energy and thickness, while the kinetic parameters may be determined from experiment-based database, such as CALPHAD~\cite{VANDEWALLE2018173} and DICTRA~\cite{AnderssonJUN}. Secondly, the number of equations to be solved is far less than the particle number. Thirdly, it is versatile and can naturally deal with morphology change, particle agglomeration/fragmentation, diffusion field, etc. Moreover, it is easily coupled with multiphysics and nonlinearity such as composition dependent mobility, strain or strain gradient, shock, anisotropic properties, etc.
The current phase-field models for sintering can be roughly classified into three catergories, based on their difference in the energy formulations.
 
\textbf{Conventional model}. The first kind is derived from the system free energy formulation~\cite{Chen1994Computer,
Chen1995Anovel,FolchPhysRevE.68.010602,Wang2006,Biswas2016}, named as "conventional model" here. The total free energy is generally written as
\begin{equation}
F=\int_\Omega \left[ f(\rho,\eta_i)+\frac{1}{2}\beta_{\rho}\lvert\nabla\rho\rvert^2 + \frac{1}{2}\beta_{\eta} \sum\limits_i \lvert\nabla\eta_i \rvert^2 \right] \text{d} v.
\end{equation}
in which $\rho$ is the conserved order parameter representing pore or substance and $\eta_i$  is the non-conserved order parameter representing grain orientation.
The kinetic equations for $\eta_i$ and $\rho$ are derived as
\begin{equation}
\dot{\eta}_i = -L \frac{\delta F}{\delta \eta_i} ,
\label{eq1stAC}
\end{equation}
and
\begin{equation}
\dot{\rho} = \nabla \cdot \left[ M \nabla \frac{\delta F}{\delta \rho} \right] ,
\label{eq1stCH}
\end{equation}
respectively. In this line of phase-field models, the temperature effect is often included in the temperature dependent $f(\rho,\eta_i)$ or model parameters, along with a heat transfer equation. However, the inclusion of non-isothermal features of sintering in this way may suffer from the thermodynamical inconsistence, and thereby the temperature gradient related kinetics are missing. Due to the simplicity, this line of models are still widely in use.

\begin{figure}[!b]
\centering
\includegraphics[width=8.4cm]{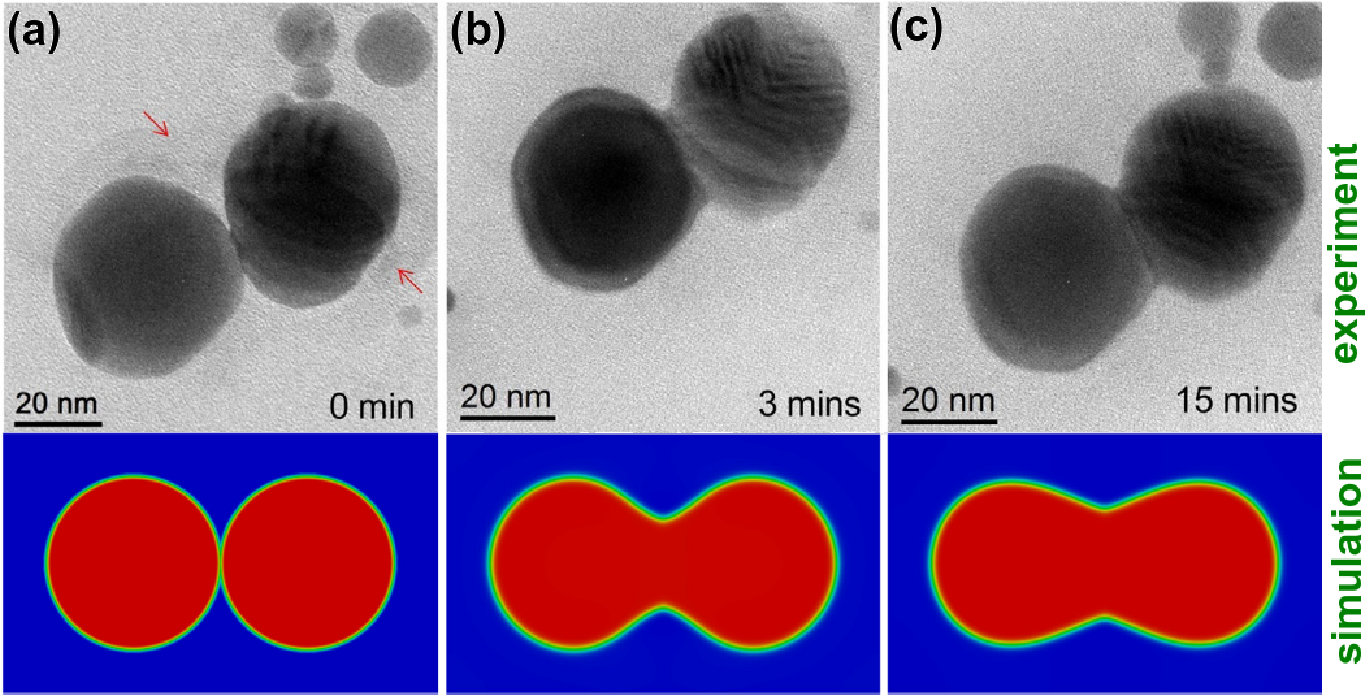}
\caption{Morphological evolution of two silver particles (diameter 40 nm) sintered at 400 $^\circ$C: comparison between experimental observation and phase-field simulation. Reproduced with permission~\cite{Chockalingam2016}. Copyright 1999, Elsevier.}
\label{figPFexp}
\end{figure}

Phase-field simulations of two-particle sintering are essential benchmarks to establish phase-field model as a successful methodology for simulating sintering process~\cite{Kumar2010,Biswas2018}. Phase-field model could incorporate elastic deformation, rigid-body motion, and heat conduction, and reveal the sintering mechanisms including neck formation and grain growth~\cite{Biswas2018}. In particular, phase-field simulations of sintering two particles with unequal size reveal three sub-processes including neck growth, coarsening accompanied with the concurrent slow grain boundary migration, and rapid grain boundary movement~\cite{Kumar2010}. The direction-dependent interface diffusion can also be considered in the phase-field model. In this way, the phenomenon that diffusion mainly occurs along the directions which are tangential to the particle surface and grain boundaries can be captured~\cite{Chockalingam2016,Deng2012}.
By using this idea, phase-field simulations are carried out to explore the sintering process of two silver nanoparticles~\cite{Chockalingam2016}. As shown in Fig.~\ref{figPFexp}, the morphological evolution of two equally sized silver nanoparticles with a diameter of 40 nm sintered at 400 $^\circ$C for 15 min is both observed experimentally and simulated by phase-field model. It can be found that there exists a good agreement between experimental observation and phase-field simulation.
In terms of the viewpoint that the solid material is characterized by a low vacancy content and the pore by a high vacancy content, a vacancy diffusion based phase-field model of sintering in two-phase and multi-phase systems is proposed \cite{Asp2006}. The model considers vacancy diffusion as the atomic mechanism for the material redistribution during sintering process. Energy formulation describing thermal vacancies in a crystal is included in the model.

\begin{figure}[!b]
\centering
\includegraphics[width=8.4cm]{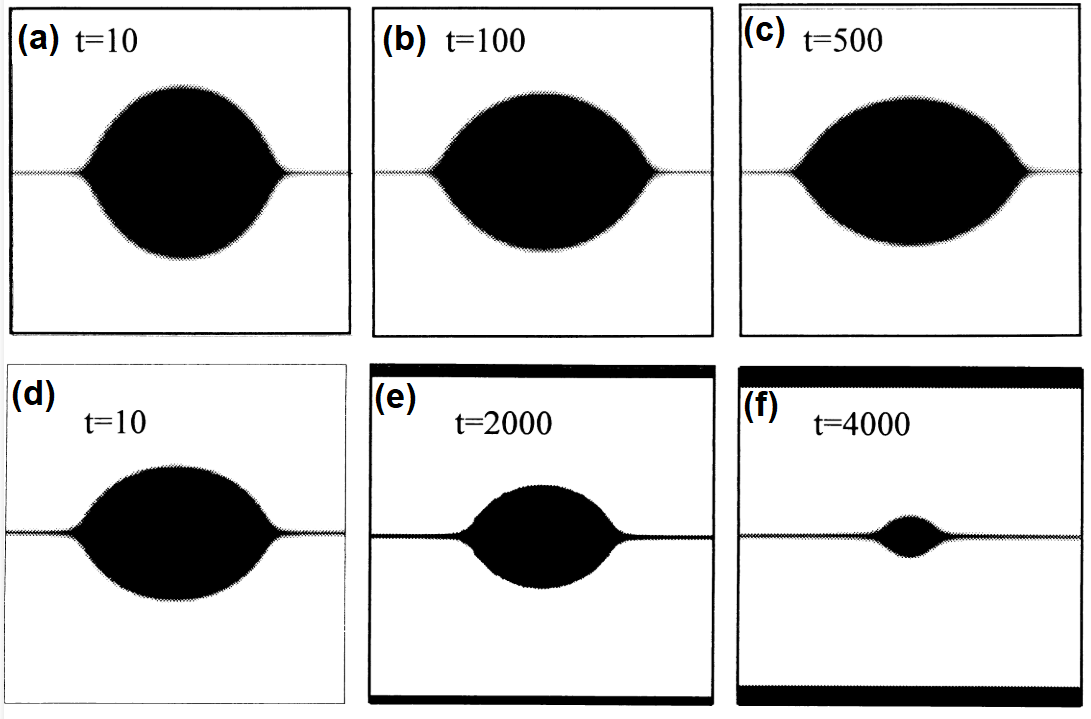}
\caption{Phase-field simulation results on the pore shape evolution during sintering: (a)--(c) with and (d)--(f) without the consideration of rigid-body motion. Reproduced with permission~\cite{Kazaryan1999}. Copyright 1999, Elsevier.}
\label{figPF2}
\end{figure}

\begin{figure}[!b]
\centering
\includegraphics[width=8.4cm]{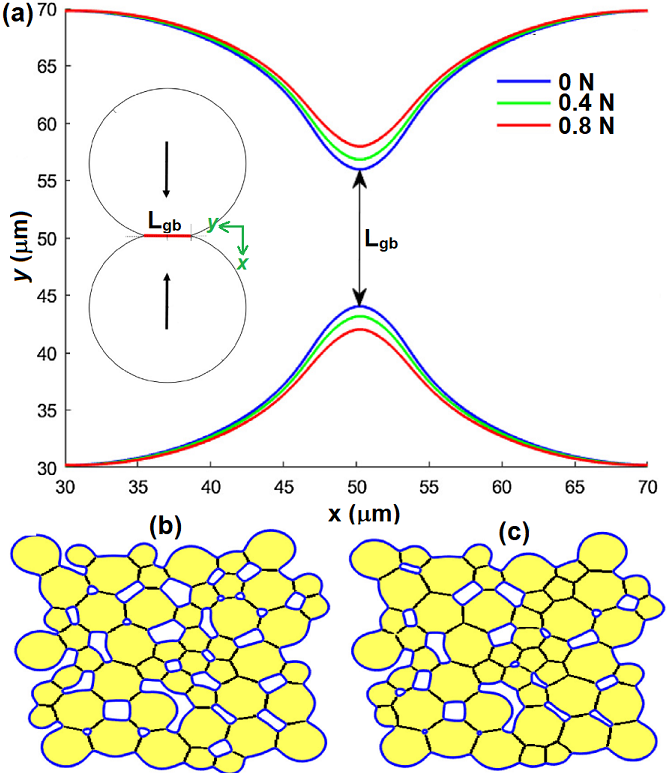}
\caption{Phase-field simulations of pressure-assisted sintering: (a) contours for two particles (with a radius of 40 $\mu$m ) sintered at different loads applied along the $x$ direction; final microstructures for 50 particles sintered at a confining pressure of (b) 0 and (c) 8 GPa. Reproduced with permission~\cite{Dzepina2019}. Copyright 2019, Elsevier.}
\label{figPF1}
\end{figure}

In the first kind of model, the rigid-body rotation and translation of grains can be described by the movement of each grain's density profile. Thus, adding an additional mass current term into Eq.~(\ref{eq1stCH}) could account for the flux density from both diffusion and rigid-body motion. Similarly, rigid-body motion also modifies the Eq.~(\ref{eq1stAC}) with an advection term~\cite{Kazaryan1999,Wang2006}.
In order to simulate the grain coalescence due to both the grain rotation and grain boundary migration, a phase-field model with multiple order parameters is proposed~\cite{Vuppuluri2019Study}. In this model, the constitutive equation for the grain rotation induced by viscous sliding and the classical phase-field model for curvature-driven grain boundary migration are coupled.
The macroscopic rigid-body motion of grains during sintering is demonstrated to remarkably influence the sintering behavior and contribute to the  densification and shrinkage of the sample , as shown in Fig.~\ref{figPF2}. The initial configuration of the two-grain system consists of equally spaced circular pores with a single flat grain boundary. It can be seen from Fig.~\ref{figPF2}(a)--(c) that in the absence of the rigid body motion, the initially circular pore relaxes to the equilibrium configuration, but the shape change does not result in shrinkage. In contrast, with the consideration of rigid-body motions, the pores begin to shrink and the sample becomes dense, as shown in Fig.~\ref{figPF2}(d)--(f)~\cite{Kazaryan1999}.
For the calculation of rigid-body movement of particles during sintering, a combination of phase-field model and DEM is proposed, which can simultaneously simulate the movement of particles and the grain growth behavior. Phase-field model is firstly utilized to evaluate the sintering forces and the contact areas in linked particles, which are then introduced into the DEM simulations for the particles' rigid-body motion~\cite{Shinagawa2014}.

In addition, the pressure-assisted sintering could be modelled by incorporating an efficient contact mechanics algorithm into the phase-field sintering model \cite{Dzepina2019}. The surface contact stress of interacting particles with arbitrary shape is considered into the model as an elastic strain energy term. Diffusive fluxes along the stress gradient could achieve the energy relaxation through deformation. Sintering behaviors due to the externally applied loads are examined, as shown in Fig.~\ref{figPF1}. The uniaxial body force in Fig.~\ref{figPF1}(a) is found to notably affect the resultant contour of sintered two particles. The length of the necks between the particles, $L_\text{gb}$, is found to increase with the external loading. The comparison of final microstructures of sintered multiple particles under different confining pressures is presented in Fig.~\ref{figPF1}(b) and (c). It is found that an 8 GPa confining pressure induces smaller and fewer pores with thicker necks between the particles.

Furthermore, an improved phase-field model with fully coupled mechanics and diffusion is proposed for sintering simulations~\cite{Zhao2022}. In order to make the normalised contact stress distribution close to that in conventional contact problems, an interpolation method for elastic modulus is adopted. Phase-field simulation results show that the stored strain energy in the contact area could accelerate the sintering process and larger pressure favors faster growth of the sintering neck~\cite{Zhao2022}.

Moreover, for simulating the microstructure evolution of hot-press sintering under non-isothermal conditions, a phase field model using the coupled thermo-mechano-diffusional equations is developed~\cite{Shulin2020}. The temperature dependent model parameters and a heat transfer equation are included in the model. It is revealed that the driving force from temperature gradient increases with the heating rates, whereas the driving forces from strain gradients and concentration gradients are identical in particles with the same shape but different heating rates~\cite{Shulin2020}.

\begin{figure}[!b]
\centering
\includegraphics[width=8.4cm]{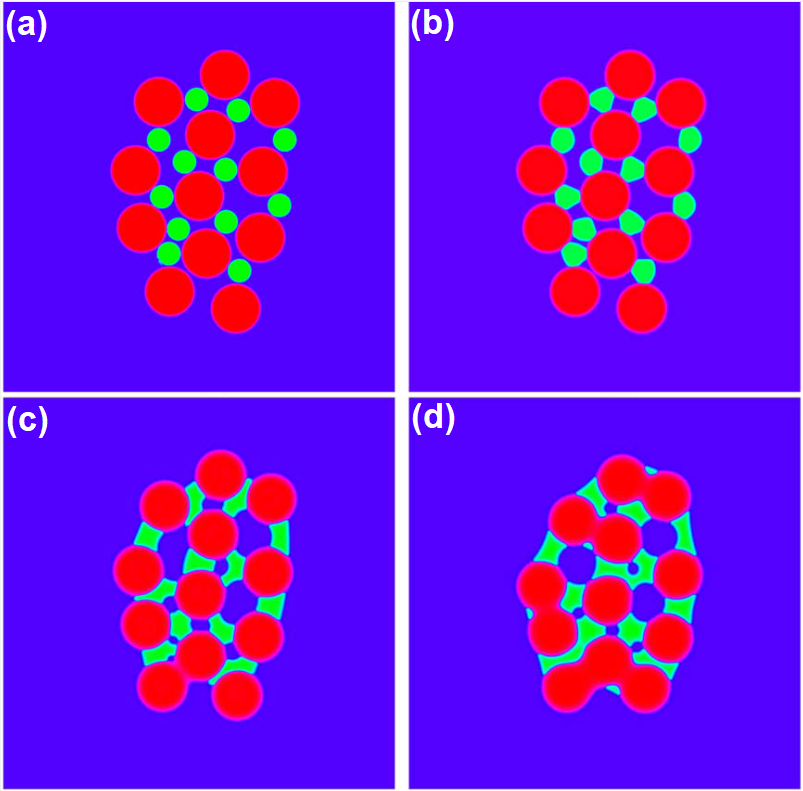}
\caption{Phase-field simulations of microstructure evolution during liquid-phase sintering with an equilibrium contact angle of 36$^\circ$ at a dimensionless time of (a) 0, (b) 2, (d) 20, and (d) 100. Reproduced with permission~\cite{Villanueva2009}. Copyright 2009, Elsevier.}
\label{figPF3}
\end{figure}

\begin{figure}[!t]
\centering
\includegraphics[width=8.4cm]{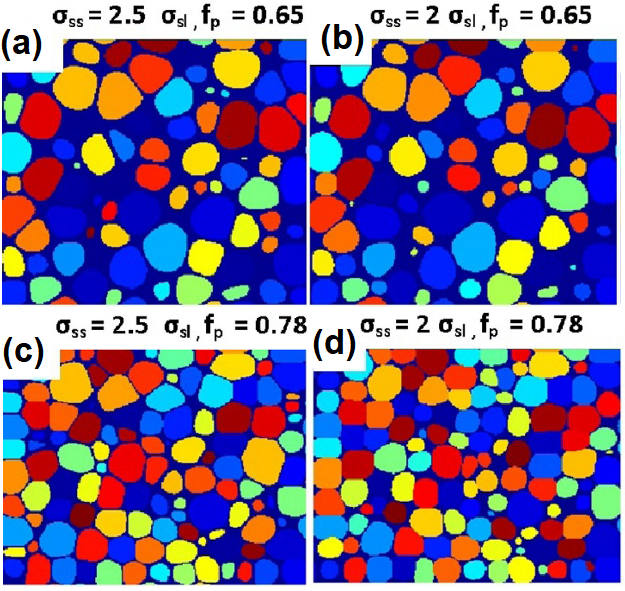}
\caption{Phase-field simulation results on the sections of the three-dimensional liquid-phase sintered microstructures with different solid volume fractions $f_\text{p}$ and solid-solid to liquid-solid interfacial energy ratios $\sigma_\text{ss}/\sigma_\text{sl}$ at a time step of 75,000. Reproduced with permission~\cite{Ravash2017}. Copyright 2017, Elsevier.}
\label{figPF4}
\end{figure}

For the case of liquid-phase sintering, a phase-field model with multicomponent and multiphase is developed~\cite{Villanueva2009}. In addition to phase-field equations in Eqs.~(\ref{eq1stAC}) and (\ref{eq1stCH}), the model consists of convective concentration and Navier--Stokes equations with surface tension forces. As an example, Fig.~\ref{figPF3} shows the microstructure evolution of a liquid-phase sintered system containing 12 liquid drops that are distributed over a matrix of 12 spherical solid particles of equal sizes. It can be seen that liquid drops rapidly wet the solid grains and two solid grains tend to contact. Due to the capillary force, the grains are rearranged and there occur more pore shrinkage/elimination and coalescence. The model is shown to be capable of capturing important dynamics in liquid-phase sintering such as rapid wetting and movement of particles owing to the capillary forces~\cite{Villanueva2009}.
For the liquid phase penetration into interparticle boundaries and the accompanied dimensional variations during the primary rearrangement stage of liquid-phase sintering, a phase-field model is designed in which the Cahn--Hilliard and Navier--Stokes equations are coupled to model the penetration of the liquid phase~\cite{MalikTahir2016}.
For the grain coarsening and grain shape accommodation in the final stage of liquid-phase sintering, a phase-field model considering a liquid phase, a polycrystalline solid phase and solid-solid/liquid interface energy is adopted~\cite{Ravash2017}. Phase-field simulations are shown to reproduce a variety of microstructural features including particle shape accommodation, Ostwald ripening, particle bonding, fully connected grain structures with liquid pockets at the grain junctions, and individual grains fully wetted by the liquid matrix. Fig.~\ref{figPF4} shows the 2-dimensional sections of the simulated microstructures of the initial system with several thousands of particles by liquid-phase sintering. It can be seen that when solid-solid to liquid-solid interfacial energy ratios $\sigma_\text{ss}/\sigma_\text{sl}=2.5$, the grains are fully wetted and a liquid network remains in the whole sample. For $\sigma_\text{ss}/\sigma_\text{sl}=2.0$, a large extend of wetting still exists and a number of particle–particle contacts also appear~\cite{Ravash2017}.
For the special case of viscous sintering, a thermodynamically consistent phase-field model is proposed, in which the conservation of mass is satisfied through the incompressibility assumption and the viscous mass flow is controlled by the Stokes equation that incorporates the surface tension~\cite{Yang2018}.


\begin{figure}[!b]
\centering
\includegraphics[width=8.4cm]{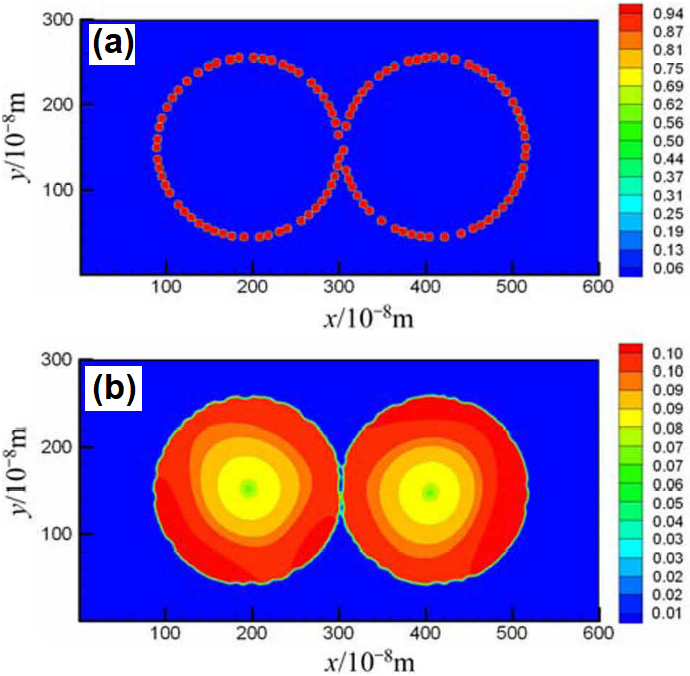}
\caption{Phase-field simulation results on the concentration field of Cu at different sintering time of Fe-Cu powders: (a) 0 s; (b) $2.7 \times 10^{-4}$ s. Reproduced with permission~\cite{Zhang2014}. Copyright 2014, Elsevier.}
\label{figPF5}
\end{figure}

For the sintering process with multiphase powders, the solute concentration is taken as the order parameter and a thermodynamic consistent phase-field model is developed~\cite{Zhang2014}. In this model, a mixture of different phases is assumed to occupy the interface region, in which the chemical potential is the same but the composition is different. The energy formulation could be taken from the thermodynamic database. The model is applied to study the sintering process of Fe-Cu powders, as shown in Fig.~\ref{figPF5}. It can be found that Cu dissolves into Fe particles step by step and finally Cu distributes uniformly~\cite{Zhang2014}.

One application of phase-field model is simulating the sintering process of ceramics~\cite{Liu2011,Wei2012}. Phase-field simulations are performed to study the microstructure evolution during the sintering of alumina-zirconia ceramics, indicating that a higher volume fraction of zirconia phase results in a smaller final grain size and a lower relative density~\cite{Wei2012}. Based on the phase-field simulation results for sintering of polycrystalline ceramics, a Gaussian process autoregressive model for capturing the microstructure evolution statistics is proposed and the reduced-order model is shown to work well~\cite{Yabansu2019}.

\begin{figure}[!b]
\centering
\includegraphics[width=8.4cm]{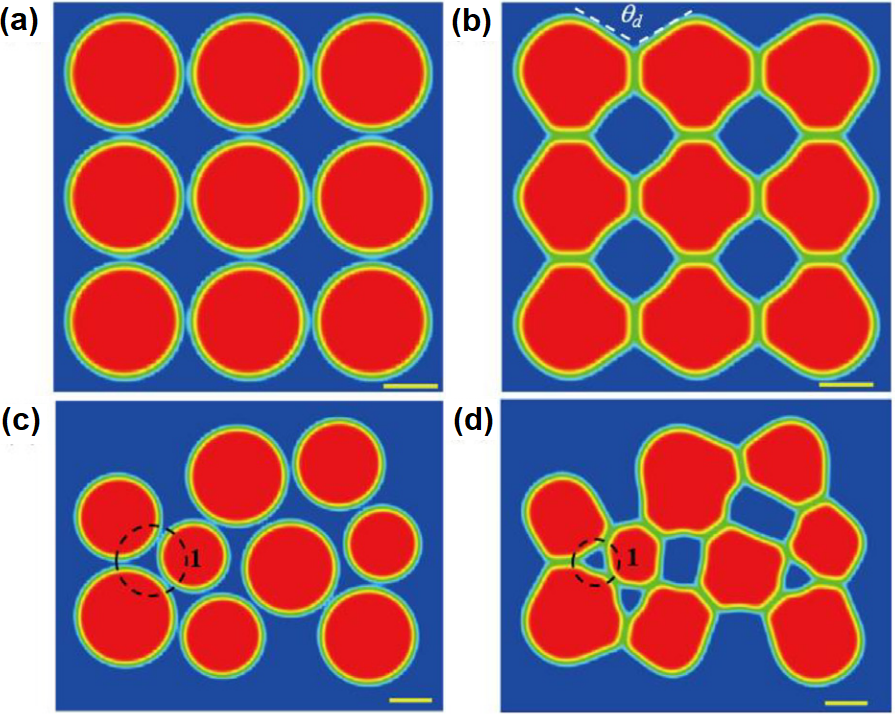}
\caption{Phase-field simulation of microstructure evolution during solid-state SLS of 316L stainless steel particles: (a) initial and (b) final configurations of nine particles with a uniform size of 40 $\mu$m); (c) initial and (d) final status of nine particles with different sizes (average size 40 $\mu$m). Reproduced with permission~\cite{Zhang2018}. Copyright 2018, Elsevier.}
\label{figPF6}
\end{figure}

Recent application of phase-field model of sintering is the simulation of sintering based additive manufacturing including SLS~\cite{Zhang2018,Zhang2020Phase} and direct metal laser sintering~\cite{Satpathy2018Direct}. Laser processing conditions are considered by integrating the phase-field model with a thermal model for the continuous heating and cooling induced by the laser irradiation. The influence of laser power and scanning speed on microstructure evolution can be investigated via phase-field simulations. Fig.~\ref{figPF6} gives the evolution of grain size and pore configuration during the solid-state SLS of 316L stainless steel particles, in which the laser power is 21 W and the scanning speed is 1 mm/s~\cite{Zhang2018}. From Fig.~\ref{figPF6}(a) and (b) with an initially uniform grain size, it can be seen that the neck size finally reaches $\sim$14 $\mu$m, the grains experience similar deformation, and large pores remain after the SLS processing. For particles with an initially non-uniform size in Fig.~\ref{figPF6}(c) and (d), the difference of particle size makes grain boundary migration toward the curvature center and the smaller particles is consumed by larger ones via volume diffusion as the SLS occurs~\cite{Zhang2018}. 
A phase-field model combining a moving heat source model is utilized to study the influence of particle features on the densities and porosities of SLS samples~\cite{Zhang2020Phase}.
Using phase-field simulations, the microstructure evolution and consolidation kinetics of Al alloys in direct metal laser sintering can be calculated~\cite{Satpathy2018Direct}. For simulation of functional oxide ceramics, where grain growth involves multiphysics like grain boundary segregation, electrostatics and mechanics, a free-energy based phase-field model has been proposed recently~\cite{Vikrant2020}.

\textbf{Grand-potential model}. The second kind is derived from the grand potential of a system~\cite{Plapp2011Unified,
Choudhury2012Grand,HOTZER2015194Large,Hotzer2019Phase,Greenquist2020}. 
In the grand-potential model, an additional non-conversed order parameter $\phi$ is introduced to represent pores ($\phi=1$) and the external void region ($\phi=0$)~\cite{KubendranAmos2020}. The conserved order parameter $c$ is introduced as the vacancy concentration. $c$ is expressed as $c=h_s c_s + (1-h_s)c_v$ with $h_s$ as a switching function that interpolates smoothly between values corresponding to the two regions. $c_s$ and $c_v$ are the vacancy concentrations in the solid and void regions, respectively.
The total grand potential of a sintering system is defined as~\cite{Greenquist2020}
\begin{equation}
\begin{split}
\Psi = \int_\Omega [ & \omega_{b}(\phi, \eta_i)+\omega_{gr}(\nabla \phi, \nabla \eta_i) + h_{s}(\phi) (f_s - c_s\frac{\mu}{V_a})  \\
&  + (1-h_s(\phi)) (f_v - c_v \frac{\mu}{V_a}) ] \text{d}v
\end{split}
\end{equation}
in which $f_s$ and $f_v$ are the Helmhotz free energy densities of the solid and void regions, respectively, $V_a$ is the atomic volume of the material, and the chemical potential of the vacancies $\mu=V_a \frac{\partial f_s}{\partial c_s} = V_a \frac{\partial f_v}{\partial c_v}$. Accordingly, the evolution equations are
\begin{equation}
\dot{\eta}_i = - L_1 \frac{\delta \Psi}{\delta \eta_i},
\end{equation} 
\begin{equation}
\dot{\phi} = - L_2 \frac{\delta \Psi}{\delta \phi},
\end{equation} 
and
\begin{equation}
\dot{\mu} =  \frac{1}{\chi} \left[ \nabla \cdot (\chi \mathbf{D} \cdot \nabla\mu) - \frac{1}{V_a} \frac{\partial c}{\partial \phi} \dot{\phi} \right],
\end{equation}
in which $\mathbf{D}$ is the diffusivity tensor and $\chi$ is the susceptibility. In this formalism, the thermodynamic energies are projected in the grand-potential space instead of the energies themselves. So the advantage is that the interface energies and the interface thickness can be decoupled.
In this way, the interface and bulk properties of the phase-field model can be adjusted independently. Therefore, the driving forces will not depend on the grid resolution, thus enabling the efficient simulation of large scale domains. Moreover, compared to other approaches, the grand-chemical-potential excess to the interface energy does not exist~\cite{Hotzer2019Phase}.

\begin{figure}[!t]
\centering
\includegraphics[width=8.4cm]{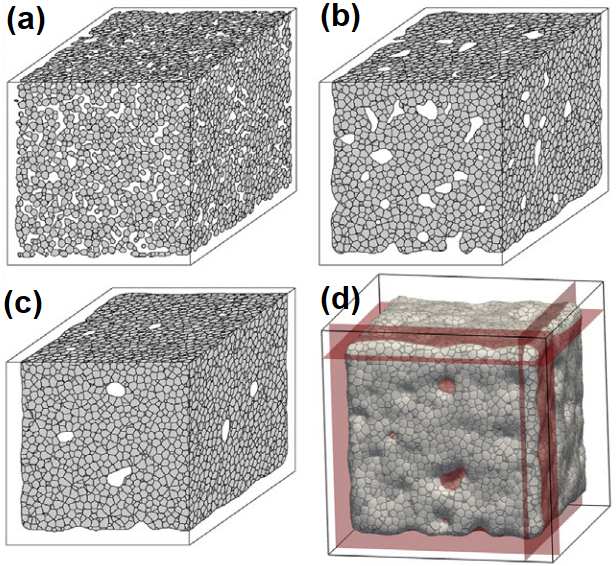}
\caption{Grand-potential phase-field simulation of microstructure evolution in 24,897 Al$_2$O$_3$-grains system sintered over time: (a) 3.65 s; (b) 68.6 s; (c) 237 s; (d) final. Reproduced with permission~\cite{Hotzer2019Phase}. Copyright 2019, Elsevier.}
\label{figPF7}
\end{figure}

\begin{figure}[!t]
\centering
\includegraphics[width=8.4cm]{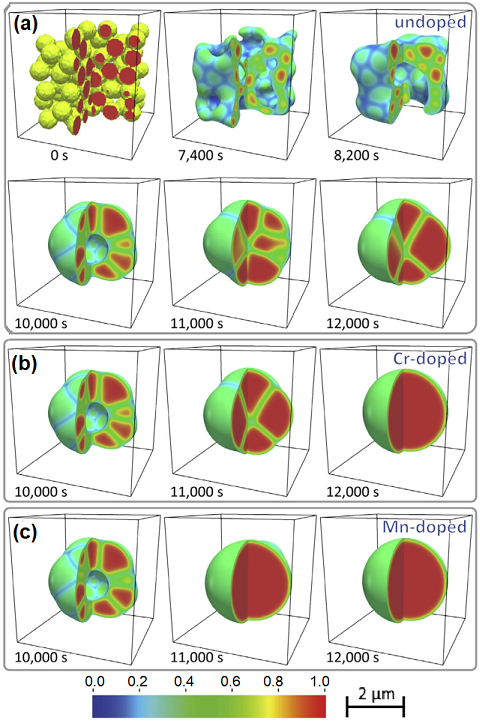}
\caption{Grand-potential phase-field simulation of microstructure evolution in 100-particle sintering: (a) pure UO$_2$; (b) Cr-doped UO$_2$; (c) Mn-doped UO$_2$. Reproduced with permission~\cite{Greenquist2020Grand}. Copyright 2018, Elsevier.}
\label{figPF8}
\end{figure}

The second-kind phase-field model has been applied to the solid state sintering of a large number of particles~\cite{Hotzer2019Phase,Greenquist2020,Greenquist2020Grand}. Grand-potential phase-field simulations considering surface, volume, and grain boundary diffusion are demonstrated to be capable of efficiently investigating the realistic green bodies with several thousands particles in three dimensions~\cite{Hotzer2019Phase}. As shown in Fig.~\ref{figPF7}, the sintering of a three dimensional green body of 24,897 Al$_2$O$_3$-grains is simulated by grand-potential phase-field model. It can be seen that starting from the loosely packed green body, the particles form sintering necks between each other and later the green body starts to densify~\cite{Hotzer2019Phase}. The second-kind model is also applied to the sintering simulations of nuclear materials such as UO$_2$ and doped UO$_2$~\cite{Greenquist2020,Greenquist2020Grand}. It is suggested that dopants have two effects on sintered UO$_2$, i.e., increasing the densification rate and the average grain size. The microstructure evolutions during the sintering of pure UO$_2$, Cr-doped UO$_2$, and Mn-doped UO$_2$ are summarized in Fig.~\ref{figPF8}. The sintering is realized by heating a powder compact from 973 K to 1973 K within 12,000 s. It can be seen that the microstructure in the initial heating stage (before 10,000 s) is identical for all three cases, since dopants do not affect sintering behaviors at low temperatures. After 10,000 s, the dopants notably accelerate the grain growth and Mn dopant achieves the fastest growth. The undoped case at 12,000 s looks very similar to the Cr-doped case at 11,000 s. However, the Mn-doped case at 11,000 s already reaches the final state, as shown in Fig.~\ref{figPF8}(c).

\begin{figure}[!t]
	\centering
	\includegraphics[width=8.4cm]{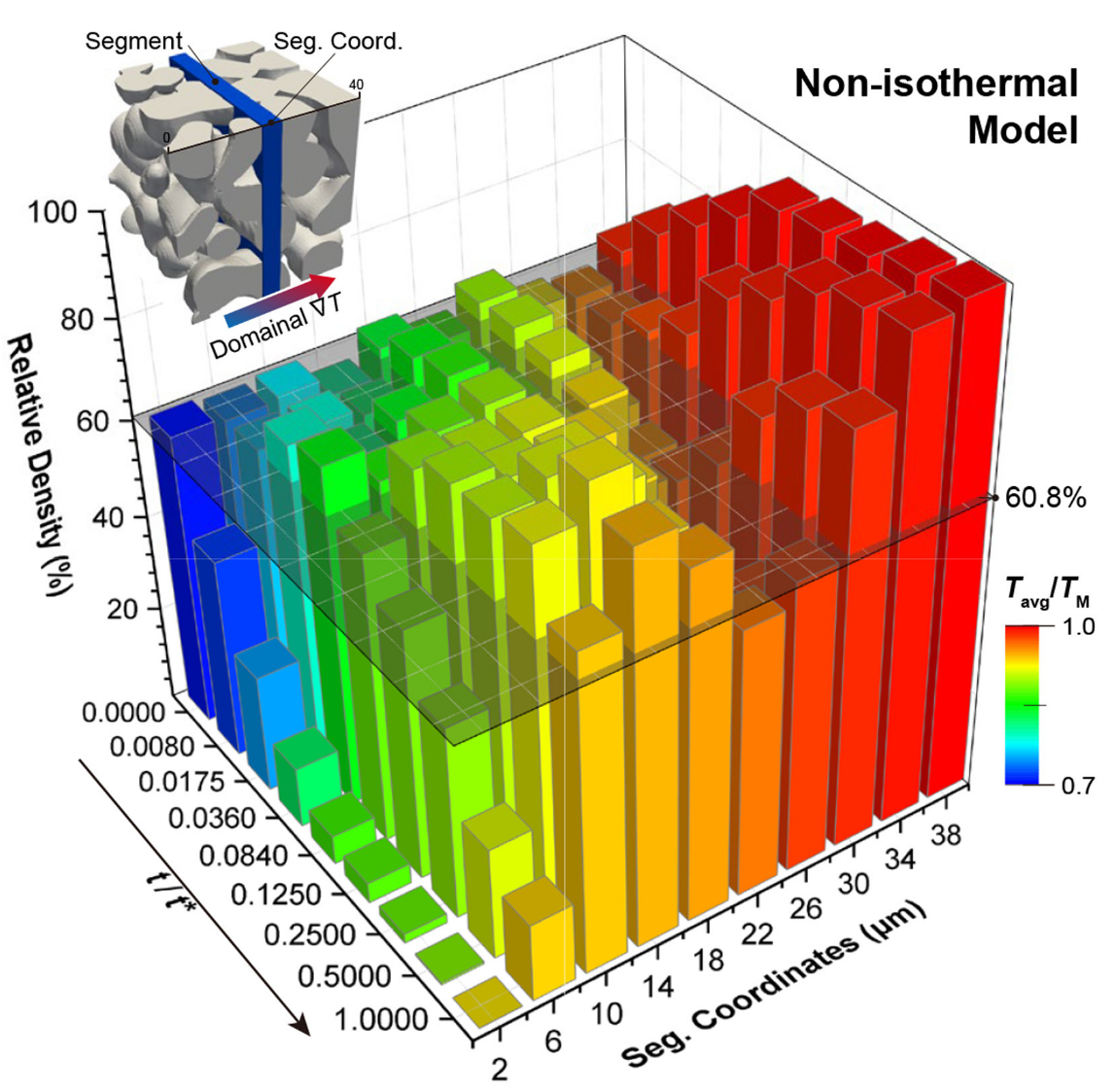}
	\caption{Non-isothermal phase-field sintering simulation results demonstrating the porosity gradient along the  $\Delta T$ direction with the color denoting the average temperature of each segment. Reproduced with permission~\cite{Yang2020Investigation}. Copyright 2020, Elsevier.}
	\label{figPFplus}
\end{figure}

\begin{figure*}[!t]
\centering
\includegraphics[width=17cm]{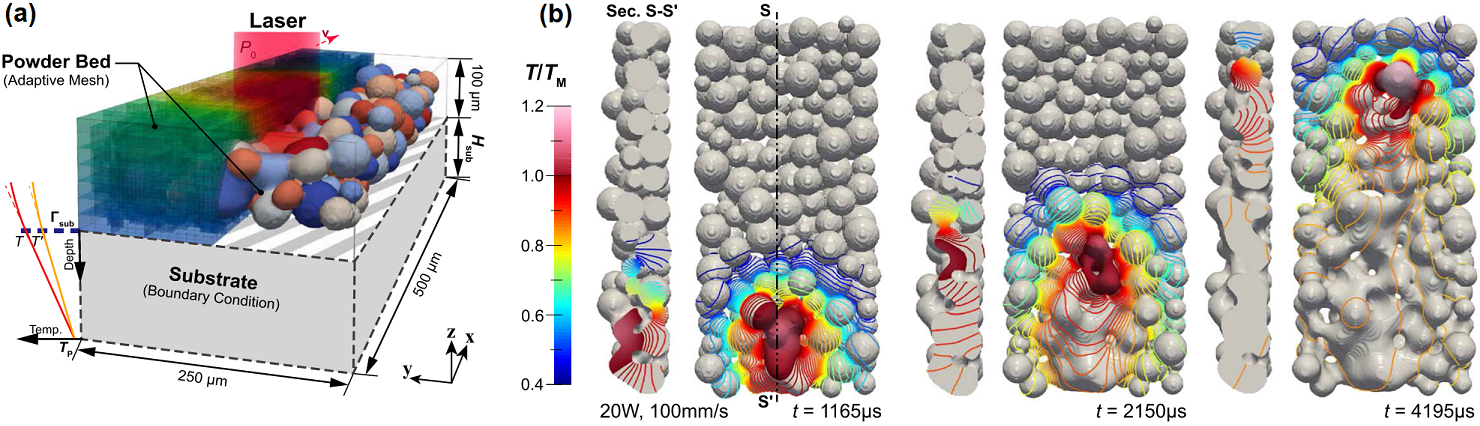}
\caption{3D non-isothermal phase-field simulation of microstructure evolution in SLS: (a) schematics of the powder bed processed by SLS; (b) simulation results on microstructure of 316L stainless steel powder bed with a laser power of 20 W and a laser scanning speed of 100 mm/s. Reproduced with permission~\cite{Yang20193Dnonisothermal}. Copyright 2019, Springer Nature.}
\label{figPF9}
\end{figure*}

\textbf{Entropy-based model.} The third kind is derived in terms of entropy formulation~\cite{Penrose1990Thermodynamically,Warren1995Prediction}, where the system entropy is expressed as
\begin{equation}
\begin{split}
& \textsl{S}(e,\rho,{\eta_{i}}) = \\
& \int_\Omega \left[ s(e,\rho,{\eta_{i}}) - \frac{1}{2}k_{\rho} \lvert \nabla \rho \rvert^2 - \frac{1}{2} k_{\eta} \sum\limits_{i} \lvert \nabla \eta_i \rvert^2 \right] \text{d} v ,
\end{split}
\label{eqEntropy}
\end{equation}
in which the entropy density $s$ is a function of internal energy density $e$, $\rho$ and $\eta_i$. The positive parameters $\kappa_\rho$ and $\kappa_\eta$ are related to the interface energy.
In this model, it is critical to construct the entropy formulation. Some special model entropy functionals and the associated kinetic equations resulting from them are constructed for phase transitions with and without a critical point or a latent heat~\cite{Penrose1990Thermodynamically}. The model is also thermodynamically consistent and is applicable to the non-isothermal case. Non-isothermal simulations using this phase-field model have been carried out to investigate the microstructure evolution in additive manufacturing~\cite{Yang20193Dnonisothermal,Yang2020Nonisothermal,2020Nonisothermal,
Wang2021Multi,Yi2021Computational}, in which the temperature gradient and high heating/cooling rate are critical.

In use of the above entropy density functional and the order-parameter interpolated internal energy formulation, Legendre transformation leads to a non-isothermal free energy functional~\cite{Yang20193Dnonisothermal,2020Nonisothermal,Yang2020Investigation}, i.e.,
\begin{equation}
\begin{split}
& \Psi\left(\rho,\eta_{i},T\right) = \\
& \int_\Omega\left[\psi \left(\rho,\eta_{i},T\right) + \frac{1}{2}T\tilde{\kappa_\rho}|\nabla\rho|^2+\frac{1}{2}T\tilde{\kappa_\eta}\sum_i|\nabla\eta_i|^2\right] \text{d} v .
\end{split}
\end{equation}  
In this way, temperature not only influences the bulk energy contributions through the temperature dependent parameters, but also modifies explicitly the gradient terms. The follow-up thermodynamic analysis leads to the fully coupled evolution equations of the modified Cahn--Hilliard and Allen--Cahn types~\cite{Yang20193Dnonisothermal,2020Nonisothermal,Yang2020Investigation} as
\begin{equation}
	\dot{\rho}=\nabla\cdot \left[ \mathbf{M} \cdot \nabla\left(\frac{\delta\Psi}{\delta\rho}\right) - \left(\frac{\delta\Psi}{\delta\rho}\right) \tilde{\mathbf{M}} \cdot \frac{\nabla T}{T} \right] ,
\end{equation} 
\begin{equation}
	\dot{\eta}_i=-L\frac{\delta\Psi}{\delta\eta_{i}} ,
\end{equation}
and the microstructure-impacted heat transfer equation as
\begin{equation}
	c_r\dot{T}+\frac{\partial e}{\partial\rho}\dot{\rho}+\sum_i\frac{\partial e}{\partial\eta_{i}}\dot{\eta_{i}} = \nabla \cdot \left[ \mathbf{k}(\rho,T) \cdot \nabla T \right] + q ,
\end{equation}
where $q$ is the heat source term and $\mathbf{k}(\rho,T)$ is the temperature- and phase-dependent heat conductivity tensor.
Diffusive mobility tensors $\mathbf{M}$ and $\tilde{\mathbf{M}}$ can be derived from the diffusivity tensor $\mathbf{D}$ according to Refs.~\cite{Schottky1965357Theory,ZHANG2012161Phase}, i.e., $\mathbf{M}=V_\text{m} \mathbf{D} / RT$ and $\tilde{\mathbf{M}}=V_\text{m} Q_\text{th}^v \mathbf{D} / (RT)^2$ with $V_\text{m}$ as
molar volume constant and $Q_\text{th}^v$ as transport heat of the vacancy.
Even though the formulations seem intricate, they cover extensively the impact of temperature and temperature gradient, including the thermocapillary, thermpphoresis, Dufour effect, etc. This makes it very attractive for simulations of sintering in a severe temperature scenario such as in the SLS with high temperature gradient and cooling rates~\cite{Yang20193Dnonisothermal,Yang2020Nonisothermal,2020Nonisothermal}. To alleviate the challenging numerical issue related to different kinetics on various time scales, the normalization of the governing equations is recommended.

Using the above model, the effect of temperature gradient on microstructure features during the unconventional sintering such as SLS and field assistant sintering is carefully examined~\cite{Yang2020Investigation}.
Fig.~\ref{figPFplus} shows the spatial and temporal relative density of different segments. It can be found that segments 34 and 38 present a chronic increasing density, while segments 2 and 6 show a chronic decreasing density. As a result, the relative density at segments 38 reaches almost 100\% while one at segment 2 is close to 0\%. Simulation results on the non-isothermal sintering of yttria-stabilized zirconia micro-particles indicate that temperature gradient induces coalescence of identical particles and there exists competition between Ficktian diffusion and thermophoresis. For the SLS, this model is extended to include the possible local surface melting and the laser-powder interaction~\cite{Yang20193Dnonisothermal}. In order to reduce the computation cost, an algorithm analogy to the minimum coloring problem is proposed so that a system of 200 grains with grain tracking algorithm can be simulated by using as low as 8 non-conserved order parameters. Fig.~\ref{figPF9} shows the typical 3D non-isothermal phase-field simulation results for the SLS of 316L stainless steel powder bed. With the powder-bed set up in Fig.~\ref{figPF9}(a), simulated microstructure and temperature evolution is give in Fig.~\ref{figPF9}(b). It can be seen that laser scanning makes particles binding together and the localized violent heating induces the partial melting of particle surfaces. The local temperature gradient around the pore region and the partial melting region is estimated up to 100 K/$\mu$m and 50 K/$\mu$m, respectively. Since the surface energy depends on temperature, such a temperature field with large gradients could induce additional mass transfer. This also makes SLS intrinsically different from the conventional isothermal sintering~\cite{Yang20193Dnonisothermal}.

\subsection{Macroscopic continuum model}
Even though the microstructural approaches above have the advantage of understanding the sintering physics, identifying the dominant densification mechanism, and predicting the microstructure evolutions, they suffer from the size restriction and are inferior in the accurate description of macroscopically effective properties of the real sintering processes. An alternative approach is the macroscopic continuum model for sintering, in which constitutive laws or analytical expressions are proposed to match the results of sintering experiments. It can be utilized to describe the real macroscopic behavior of sintering process by adjusting parameters based on the experimental data without the consideration of parameters' physical meaning. It is also applicable to the design of sintering processing parameters for obtaining a sintered part with the desired density and shape.

\begin{figure}[!t]
\centering
\includegraphics[width=8.4cm]{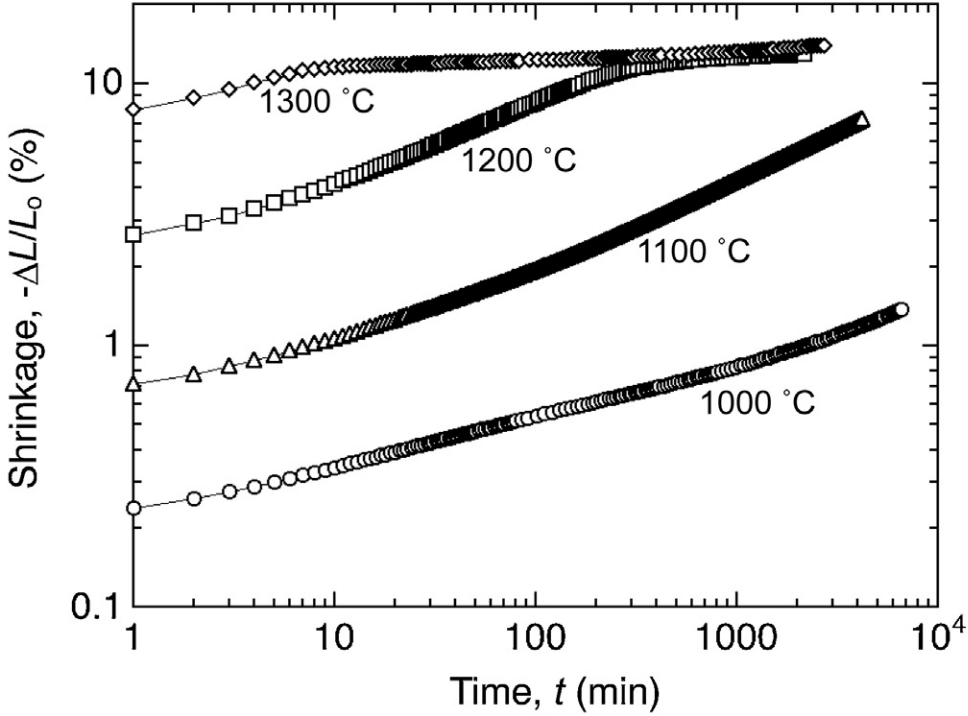}
\caption{Shrinkage during the isothermal sintering of 8YSZ at different temperatures. Reproduced with permission~\cite{Kim2016Densification}. Copyright 2016, Elsevier.}
\label{figMCM1}
\end{figure}

The first kind is the \textbf{sintering kinetics model} that could capture the densification process. By using the experimental shrinkage data, a formal kinetics approach is proposed by Palmour in the 1980s~\cite{Palmour1987Rate}.
At the initial-stage sintering, the most dominant phenomenon is the rapid growth of interparticle neck. A two-sphere model is widely used to analyze the neck growth with an initial shrinkage of 3--5\%~\cite{Kingery1955Study, Johnson1963Diffusion}. In this two-sphere model, the densification kinetics at the initial sintering stage could be described by~\cite{Kingery1955Study, Johnson1963Diffusion}
\begin{equation}
\frac{\Delta L}{L_0} = B^m (t-t_0)^m
\label{ski1}
\end{equation}
in which $\Delta L$ and $L_0$ is the length change and the initial length of the sintered specimen, respectively, $B$ is a model constant, and $t_0$ is the time when the neck starts to grow. $m$ is the parameter related to the densification mechanism, with $m=1/2$ for lattice diffusion and $m=1/3$ for grain-boundary diffusion. However, the two-sphere model in Eq.~(\ref{ski1}) only works well for the case that shrinkage occurs in one dimension. It has limitations when applied to the real sintering case, since the actual powder compacts possess a microstructure of complicated particle arrangements in three dimensions. For instance, the shrinkage behavior during the isothermal sintering of 8YSZ in Fig.~\ref{figMCM1} gives the value of $m$ around 0.14--0.21 at 1,000--1,200 $^\circ$C, which are much lower than 1/2 or 1/3 in the two-sphere model. This deviation could be ascribed to the wide-range distribution of the number of neighboring particles and the instant size of interparticle neck in the real compacts. This indicates that a many-particle model is required to describe more explicitly the densification kinetics at the initial stage.

At the intermediate stage, a typical rate equation of densification is expressed as~\cite{Wang1990Estimate}
\begin{equation}
\dot{\rho} = \frac{A}{T} \text{exp}\left(\frac{-Q}{RT}\right) \frac{ f(\rho)}{d^n}
\end{equation}
in which $\dot{\rho}$ is the densification rate, $T$ the absolute temperature, $Q$ the activation energy, $n$ an exponent of the grain size $d$, $R$ the gas constant, and $f(\rho)$ a function of the relative density $\rho$. From the collection of experimental data at a fixed density, $Q$ and $n$ can be estimated from the $\text{ln}(T\dot{\rho})$ \textit{vs} $1/T$ slope and $\text{ln}\dot{\rho}$ \textit{vs} $\text{ln}d$ slope at a fixed density, respectively. The grain size is not significantly different in the pressureless sintering, but it could be controlled at a fixed density during pressure-assisted sintering. At the intermediate stage, the grain growth is insignificant.

\begin{figure*}[!h]
\centering
\includegraphics[width=13cm]{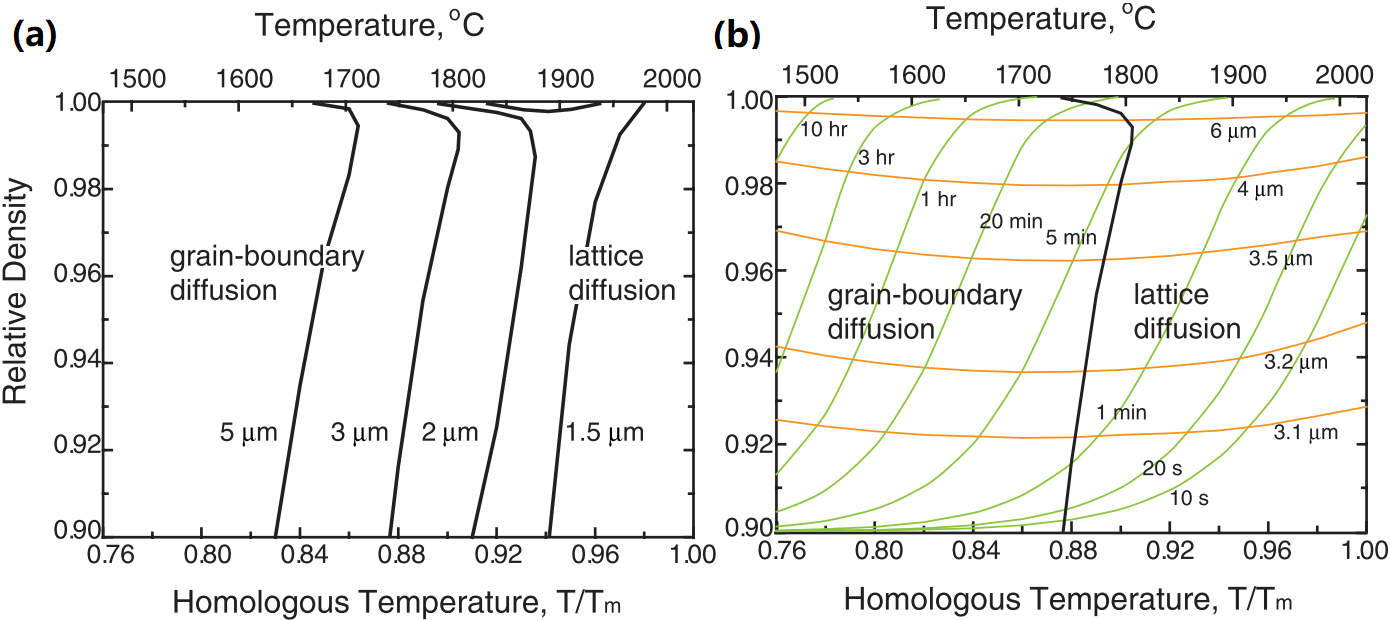}
\caption{Sintering diagrams of alumina at the final-stage sintering. (a) Relative density for various grain sizes at the beginning of the calculation (90\% relative density). (b) Sintering kinetics of a powder compact at a constant grain size and at a constant time (grain size 3 $\mu$m). Reproduced with permission~\cite{Kang2004Sintering}. Copyright 2004, Elsevier.}
\label{figMCM2}
\end{figure*}

At the final stage of solid state sintering, the densification is characterized by the shrinkage of isolated pore located at grain corners and junctions. Coble~\cite{Coble1961Sintering1,Coble1961Sintering2} proposed a geometrical model for the final-stage sintering of bcc-packed tetrakaidecahedral grains with spherical pores at 24 corners of a grain. Coble defines the concentric spherical lattice diffusion of atoms from a distance of $r_2$ to the surface of the pore with a radius of $r_1$. If $r_1 \ll r_2 $, the densification kinetics can be expressed as~\cite{Coble1961Sintering1,Coble1961Sintering2}
\begin{equation}
\dot{\rho} = \frac{288D_1 \gamma_\text{s} V_\text{m}}{RTd^3}
\label{skf1}
\end{equation}
in which $D_1$ is the lattice diffusion coefficient, $\gamma_\text{s}$ the specific surface energy, and $V_\text{m}$ the molar volume. Coble’s model in Eq. (\ref{skf1}) has been widely used in the interpretation and prediction of the densification at the final-stage sintering governed by lattice diffusion.
Since the densification rate is found proportional to the pore size, an alternative model for the lattice diffusion is proposed as~\cite{Kang2004Sintering}
\begin{equation}
\dot{\rho} = \frac{441 D_1 \gamma_\text{s} V_\text{m}}{RTd^3} (1-\rho)^\frac{1}{3}.
\label{skf2}
\end{equation}
Moreover, in order to consider the grain boundary as an atom source for densification, the model for the grain-boundary diffusion is expressed as~\cite{Kang2004Sintering}
\begin{equation}
\dot{\rho} = \frac{733 D_\text{b} \delta_\text{b} \gamma_\text{s} V_\text{m}}{RTd^4}
\label{skf3}
\end{equation}
in which $D_\text{b}$ is the grain boundary diffusion coefficient and $\delta_\text{b}$ is the diffusion thickness of grain-boundary diffusion. In contrast to Coble's model, the models in Eqs.~(\ref{skf2}) and (\ref{skf3}) take into account the role of grain boundaries and the diffusion area in densification. Fig.~\ref{figMCM2} shows two final-stage sintering diagrams of alumina by using Eqs.~(\ref{skf2}) and (\ref{skf3})~\cite{Kang2004Sintering}. It can be found from Fig.~\ref{figMCM2}(a) that the densification occurs by grain boundary diffusion at a homologous temperature lower than 0.85 and the lattice diffusion dominates the densification at high temperatures. The dominant mechanism could change from the lattice diffusion to the grain boundary diffusion when sintering continues at high temperatures, owing to that the pore size reduces with the increasing densification. This indicates that the densification mechanism can vary with the change of pore size during sintering.

\begin{figure}[!b]
\centering
\includegraphics[width=8.4cm]{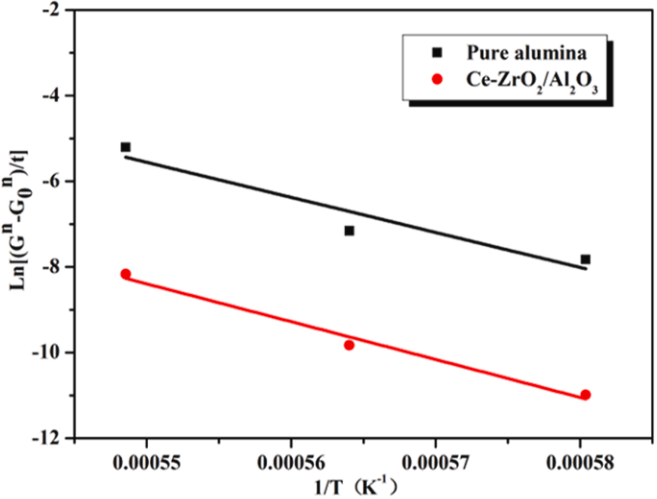}
\caption{Arrhenius plots for the sintering of undoped Al$_2$O$_3$ and Ce-ZrO$_2$/Al$_2$O$_3$ ceramics. Reproduced with permission~\cite{Wu2020Sintering}. Copyright 2020, Elsevier.}
\label{figMCM3}
\end{figure}

The grain growth at the final-stage sintering is thought to be controlled by the pore drag with the assumption that the pores are monosized, i.e.,~\cite{Zhao1992Effect,Wu2020Sintering}
\begin{equation}
\dot{d} = \frac{C D N_\text{g}^q } {(1-\rho)^m d^n}
\end{equation}
in which $C$ is a constant depending on sintering temperature and grain growth mechanism,  $N_\text{g}$ the number of pores per grain, and $q$, $n$, and $m$ are exponents related to the controlling mechanism. For the surface diffusion controlled pore drag, i.e., $D$ is the surface diffusion coefficient, $q = 1/3$, $n = 3$, $m = 4/3$. For the lattice diffusion controlled pore drag ($D$ is the lattice diffusion coefficient) and the vapor phase diffusion controlled pore drag ($D$ is the vapor phase diffusion coefficient), $q = 0$, $n = 2$, $m = 1$. For the evaporation or condensation controlled pore drag, i.e., $D$ is the sticking coefficient, $q = -1/3$, $n = 1$, $m =-2/3$~\cite{Zhao1992Effect}. With the experimental time \textit{vs} grain size data at hand, a cubic or fourth-order kinetic relationship by fitting data also works for the sintering of undoped alumina and CeO$_2$/ZrO$_2$ co-doped alumina~\cite{Wu2020Sintering}.
In order to obtain the activation energy that controls the grain growth process, the following equation~\cite{Senda1990Grain}
\begin{equation}
d^n - d_0^n = t K_0 \text{exp}\left(-\frac{Q}{RT}\right)
\label{skf12}
\end{equation}
is widely used, in which $K_0$ is a constant, $Q$ the apparent activation energy, $d$ the average grain size at time $t$, $d_0$ the initial grain size, and $n$ the kinetic grain growth exponent. Based on Eq.~(\ref{skf12}), the slope of an Arrhenius plot of ln[$(d^n-d_0^n)/t$] \textit{vs} the inverse temperature $1/T$ could give the activation energy controlling the grain growth process for an isochronal series of composites~\cite{SPEIGHT1968Growth,ARDELL1972On,Xiao2021,Wu2020Sintering}. For instance, the Arrhenius plots in Fig.~\ref{figMCM3} indicate the activation energies for the undoped Al$_2$O$_3$ and Ce-ZrO$_2$/Al$_2$O$_3$ sintering samples are 680.2 and 734.8 kJ/mol, respectively. The higher activation energy of sintering Ce-ZrO$_2$/Al$_2$O$_3$ ceramics could be attributed to the enthalpy for defect formation and/or the liquid phase at grain interfaces \cite{Wu2020Sintering}.

The second kind is the \textbf{continuum mechanics model}, which is a phenomenologically macroscopic model and could be able to predict the evolution of component shape under the influence of macroscopic parameters during sintering in addition to the sintering kinetics~\cite{Olevsky1998Theory,Riedel1994EquilibriumPore}. The model is originated from the continuum mechanical theory of plastic deformation of porous bodies~\cite{GREEN1972Aplasticity,SHIMA1976285Plasticity}.
All the continuum mechanics sintering models include a set of constitutive equations that correlate the shrinkage of a porous body with the viscous parameters and the generalized state of stress. The sintered porous body can be considered as a visco-plastic continuum. For the case of isotropic sintering~\cite{Riedel1994Densification,Jagota1990Isotropic, Rojek2017,Petersson2007Modelling,Kim2003,Baranov2015Phenomenological}, in terms of the principal stresses $\sigma_I$ and strain ${\varepsilon}_I$, one has \cite{Bordia1988On1,Bordia1988On2,Bordia2006Anisotropic}
\begin{equation}
\dot{\varepsilon}_I = \dot{\varepsilon}_I^\text{s} + \frac{1}{E_\text{vis}} \left[ \sigma_I - \nu_\text{vis} \left( 3\sigma^\text{m} - \sigma_I \right) \right]
\label{sm1}
\end{equation}
in which $I=1,2,3$ is the three principal coordinate directions,  $\dot{\varepsilon}_I^\text{s}$ the sintering induced free strain rate, $\sigma^\text{m}$ the mean (or hydrostatic) stress, and $E_\text{vis}$ and $\nu_\text{vis}$ the uniaxial viscosity and the viscous Poisson's ratio of the porous sintering body, respectively. Accordingly, the densification rate can be expressed by the trace of the strain rate tensor~\cite{Riedel2005}, i.e.,
\begin{equation}
\dot{\rho} = - \rho (\dot{\varepsilon}_1 + \dot{\varepsilon}_2 + \dot{\varepsilon}_3) 
\label{sm2}
\end{equation}
in which $\dot{\varepsilon}_I ~ (I=1,2,3)$ is the strain along the three principal coordinate directions.
The intrinsic sintering stress $\sigma^\text{s}_I$ or sintering potential as the driving force for sintering due to the interfacial energies of pores and grain boundaries can be derived as
\begin{equation}
\sigma^\text{s}_I = E_\text{vis} \dot{\varepsilon}_I^\text{s}.
\end{equation}
The grain growth rate is described by the modified classical Hillert law~\cite{Riedel2005}, i.e., 
\begin{equation}
\dot{d} = \frac{\gamma_\text{b} M_\text{b}}{2d} \frac{F_\text{d}}{F_\text{p}}
\label{sm3}
\end{equation}
in which $M_\text{b}$ is the grain boundary mobility exhibiting an Arrhenius-type temperature dependence, and $\gamma_\text{b}$ the specific energies of grain boundary. The factor $F_\text{d}$ is a function of $d$ and is introduced to consider that the powder usually does not have the steady-state grain size distribution. The factor $F_\text{p}$ represents the drag that pores exert on migrating grain boundaries. Some typical formulations for $F_\text{d}$ and $F_\text{p}$ are given in~\cite{Riedel2005}. Eqs.~(\ref{sm1}) to (\ref{sm3}) define the solid state sintering model and give the evolution equations for the strain rate, relative density, and grain size.

\begin{figure*}[!ht]
\centering
\includegraphics[width=14.5cm]{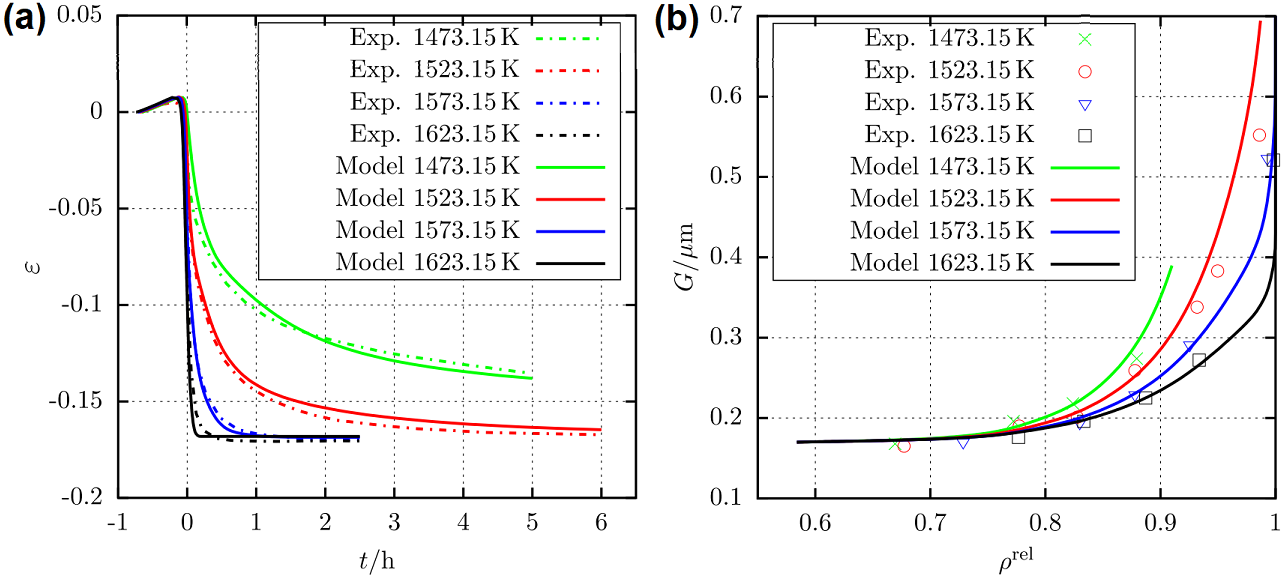}
\caption{Comparison between model (including viscous and thermo-hyperelastic effects) and experimental data for free sintering experiments: (a) logarithmic strain $\varepsilon$ \textit{vs} time for different holding temperatures; (b) grain size \textit{vs} relative density for different holding temperatures. Reproduced with permission~\cite{Stark2018Acontinuum}. Copyright 2018, Elsevier. Experimental data reproduced with permission~\cite{Zuo2004Temperature,Aulbach2004Laser}. Copyright 2004, Elsevier; 2004, Springer.}
\label{figMCM4}
\end{figure*}

The determination of the above constitutive sintering parameters such as the viscosities, the viscous Poisson's ratio, and the intrinsic sintering stress is nontrivial and usually relies on the design of experimental measurements. A discontinuous hot forging approach is employed to extract the temperature-dependent constitutive sintering parameters of alumina~\cite{Zuo2004Temperature}. With this approach, uniaxial viscosity and sintering stress, as well as uniaxial and bulk viscosities, as functions of density and temperature for an isotropic microstructure could be experimentally obtained. Using the intermittent loading method for WC-Co powder compacts~\cite{GILLIA2001Viscosity}, the free sintering strain rate part and the viscoplastic strain rate part of the constitutive model can be determined~\cite{Kim2003}. From the viscoplastic strain rate part, the axial viscosity and the viscous Poisson's ratio can be determined.

The isotropic model can be extended to the sintering of tiny glass beads that are modeled as isotropic elasto-plastic solid by using a bilinear stress-strain constitutive relationship~\cite{Fang2016Sintering}. In this model, the coupled mechanical and thermal response of the glass beads under cyclic compressive loadings can be revealed. The gradients of temperature and stress from the interiors toward the contact points are found in the triaxial compression experiments of tiny glass beads. The model can also be generalised to the sintering of materials with bimodal pore distribution~\cite{Kuzmov2005Mechanics} and phase transformations~\cite{Olevsky1997Densification}. The model can further be validated by molecular dynamics simulation for the sintering of straight-chain aggregate~\cite{Hawa2007Molecular} and nanoparticle assembly~\cite{Zachariah1999Molecular}.

\begin{figure*}[!h]
\centering
\includegraphics[width=15cm]{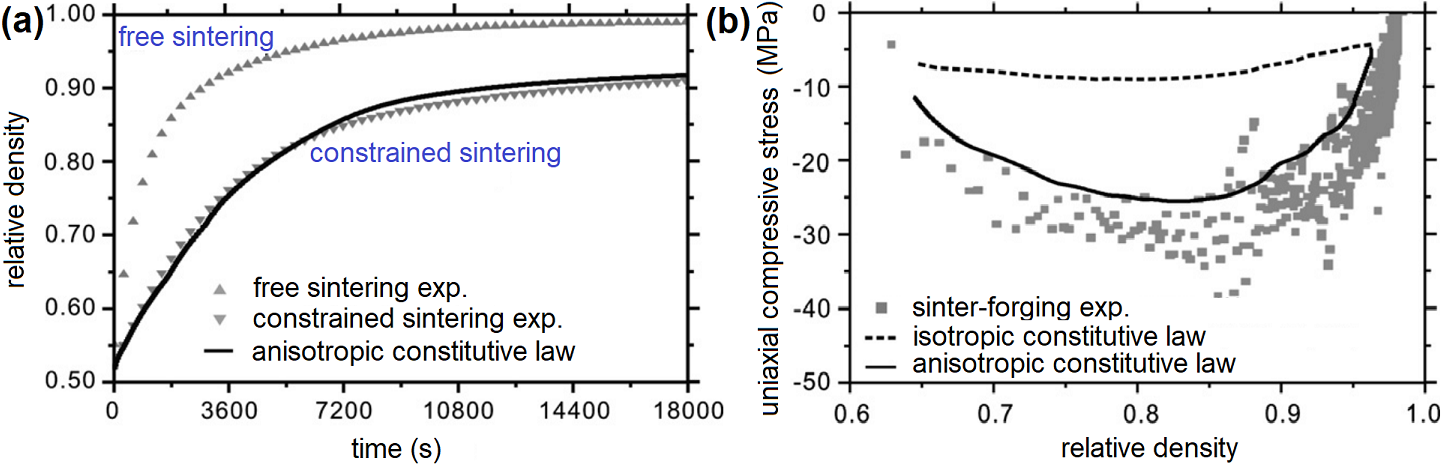}
\caption{(a) Densification curves for the free and constrained sintering. (b) Compressive stress as a function of relative density in the case of zero lateral shrinkage in sinter-forging. The calculation results from isotropic and anisotropic constitutive laws are compared with the experimental data. Reproduced with permission~\cite{Li2010Predicting}. Copyright 2010, Elsevier. The isotropic calculation results and experimental data reproduced with permission~\cite{Bordia2006Anisotropic}. Copyright 2006, Elsevier.}
\label{figMCM5}
\end{figure*}

\begin{figure*}[!t]
\centering
\includegraphics[width=11.5cm]{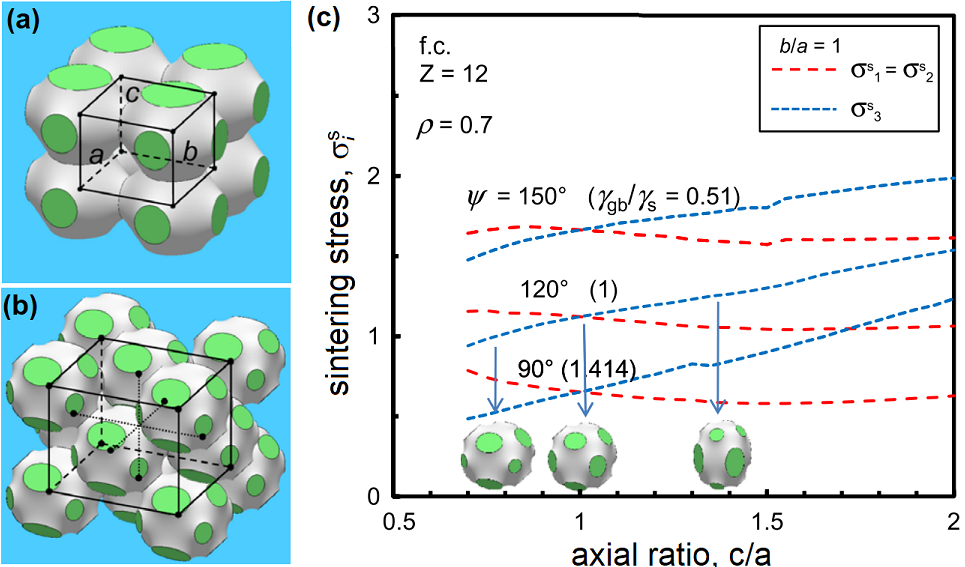}
\caption{Local arrangements of particles in orthotropic volume elements ($c/a = 0.8$): (a) simple tetragonal lattice, (coordination number $Z=6$, $b/a =1$, relative density $\rho = 0.6$) ; (b) face-centered orthorhombic lattice ($Z= 12$, $b/a = 1.1$, $\rho = 0.74$). (c) Sintering stress components in face-centered structure as a function of the axial ratio $c/a$. Reproduced with permission~\cite{Wakai2009Anisotropic}. Copyright 2009, Elsevier.}
\label{figMCM6}
\end{figure*}

Recently, starting form the isotropic model, a phenomenological thermodynamically consistent finite deformation sintering model incorporating viscous as well as thermo-hyperelastic effects and the associated finite element implementation is proposed~\cite{Stark2018Acontinuum}. It is found that the model is capable of describing the sintering process and predicting the after-sintering residual stresses of alumina ceramics under the assumption of an isotropic microstructure. The model is applied to simulate the free sintering case that is originally explored by experiments~\cite{Zuo2004Temperature, Aulbach2004Laser}. In detail, a sample of alumina is first heated from room temperature to the holding temperature $T^\text{max}$ with a heating rate of 30 K/min. Then the temperature is held constant for up to 6 h, and axial and radial strains are recorded. The simulation and experimental results are compared in Fig.~\ref{figMCM4}, and the excellent agreement between simulation and experiment is achieved.

However, the above isotropic constitutive laws are unable to describe the sintering process of thin constrained films, co-sintering of multi-layered systems, and sinter-forging. In a constrained sintering of alumina thin films, it is found that the comparison of experimental densification behavior with the isotropic continuum mechanics model predictions highlights the inadequacy of the isotropic models~\cite{Guillon2007Constrained}. This implies the necessity of a new continuum formulation accounting for the intrinsic anisotropy originated from the sample manufacturing and the extrinsic anisotropy due to sintering under non-hydrostatic external stresses or constrained sintering. A general transversely isotropic viscous formulation is developed for the cases of constrained densification of films and sinter forging, in which five constitutive parameters and two free densification rates are needed~\cite{Bordia2006Anisotropic}. Moreover, an anisotropic constitutive law with the state of the material described by the sintering strains rather than the relative density is developed~\cite{Li2010Predicting}. For the free sintering, the anisotropic constitutive law reduces to a conventional isotropic one. As shown in Fig.~\ref{figMCM5}(a), the anisotropic constitutive law can readily reproduce the densification behavior of constrained sintering. Fig.~\ref{figMCM5}(b) indicates that an isotropic constitutive law cannot correctly calculate the uniaxial compressive stress that is required to achieve zero radial shrinkage in the sinter-forging experiment. On the contrary, the anisotropic constitutive law yields results agreeing well with the experimental data~\cite{Li2010Predicting}.

In the case of anisotropic shrinkage in the many-particle sintering process, the sintering stress, i.e., thermodynamic driving force for the anisotropic shrinkage, is also found to be anisotropic~\cite{Wakai2009Anisotropic}. The sintering stress tensor for sintering of particles that are arranged in orthotropic symmetry is calculated scrupulously by the force balance method, the energy method, and the volume averaging method. The anisotropic packing structure is modelled by local arrangements of identical particles in rhombic volume elements, such as simple tetragonal lattice in Fig.~\ref{figMCM6}(a) and face-centered orthorhombic lattice in Fig.~\ref{figMCM6}(b). The calculated sintering stress components $\sigma^\text{s}_1$ and $\sigma^\text{s}_3$ for the tetragonal structures are found to depend on the axial ratio $c/a$, as shown the face-centered case in Fig.~\ref{figMCM6}(b). When $c/a$ exceeds 1, $\sigma^\text{s}_3$ is larger than $\sigma^\text{s}_3$ and thus the deviatoric sintering stress makes the elongated grains deform and more isotropic in the face-centered lattice (Fig.~\ref{figMCM6}(b)). Anisotropic shrinkage is revealed to be driven by the deviatoric component of the sintering stress tensor, which arises during the sintering of non-isotropic packing of flat or elongated particles~\cite{Wakai2009Anisotropic}.

\begin{figure}[!ht]
\centering
\includegraphics[width=8cm]{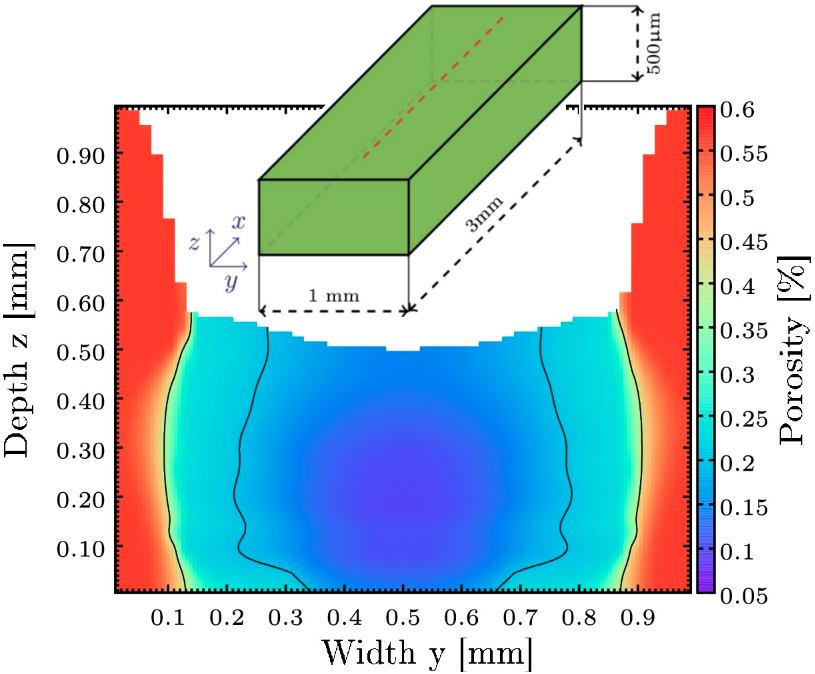}
\caption{
Specimen geometry for SLS PA12 and simulated porosity distribution in the $y$-$z$ section at $t=2.6$ s. Reproduced with permission~\cite{Mokrane2018Process}. Copyright 2018, Elsevier.
}
\label{figpolymerSLS2}
\end{figure}

\begin{figure}[!t]
\centering
\includegraphics[width=8.4cm]{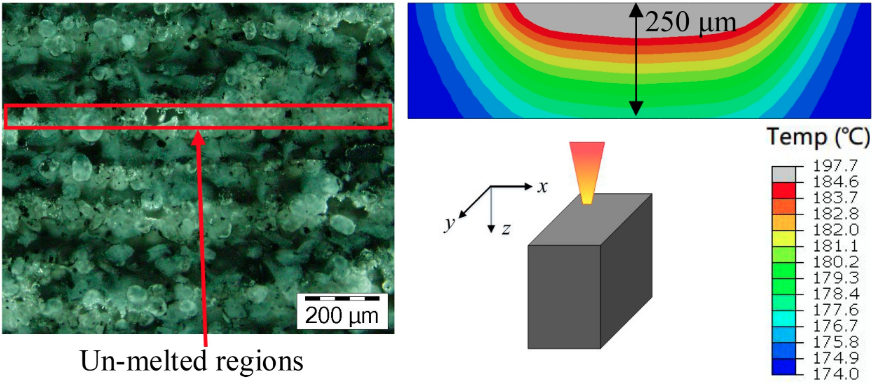}
\caption{
Experimental microstructures and simulated fusion zone after the scanning of three tracks during SLS CNTs/PA12 composite. Reproduced with permission~\cite{Shen2018Development}. Copyright 2018, Elsevier.
}
\label{figpolymerSLS1}
\end{figure}

For the \textbf{SLS of polymer}, most of the available modelling and simulation methodologies belong to the continuum model at macroscale~\cite{Mokrane2018Process,Brighenti2021Laser-based,Papazoglou2021}. 
The most important difference between sintering of polymeric and metal/ceramic powders is the densification of the molten mass after the powder-particle coalescence due to the melting or glass transition of polymers~\cite{Hornsby1992Mechanism}. Above the glass transition or melting temperatures, the viscous flow remains Newtonian if the sintering shear rates are extremely low.
Frenkel~\cite{frenkel1945viscous} considered the coalescence of two equally sized incompressible spherical particles and derived a fundamental law for the neck radius as a function of time. Alternatively, Scherer~\cite{SCHERER1977Sintering} used cubic cells containing intersecting cylinders with radius as average particle size to approximate porous material, and obtained cylinders' aspect ratio as a function of time. Subsequently, Pokluda et al.~\cite{Pokluda1997Modification}, Bellehumeur et al.~\cite{Bellehumeur1998therole}, and Balemans et al.~\cite{Balemans2017Sintering} modified Frenkel's model to consider sintering angle evolution, viscoelastic behavior of polymers, and more complex viscoelastic models, respectively. However, for SLS processing of a large number of polymer particles, these two-particle models are powerless. In contrast, researchers usually do not consider the individual polymer particles, and take the particles assemble as a continuum medium and a continuum view of heat conduction in the powder bed. For instance, Andena et al. ~\cite{Andena2004Simulation} utilized a finite element model to simulate the sintering process of polytetrafluoroethylene (PTFE) cylinders by accounting for three separate stages: thermal, deformation, and stress analysis.
Peyre et al.~\cite{Peyre2015Experimental} also performed finite-element thermal simulation to study the thermal cycles and fusion depths during SLS of two polymers: polyamide (PA12) and polyetherketoneketone (PEKK).
For SLS of crystalline (e.g., nylon-12) and composite crystalline (e.g., glass-filled nylon-11) polymers, Childs et al.~\cite{Childs2001Selective}  included latent heat of melting, viscous sintering law of crystalline material, and finite depth of laser absorption in the thermal finite element simulation.
In addition, Mokrane et al.~\cite{Mokrane2018Process} simulated SLS of semi-crystalline PA12 powders by using finite-volume method for space discretization and second-order semi-implicit Crank--Nicolson scheme for time discretization. They took PA12 powder bed as a homogeneous medium and formulated the multiphysical models of laser power distribution, thermo-physical properties, coalescence, gas diffusion, fusion, and crystallization. The simulated porosity distribution and densification are shown in Fig.~\ref{figpolymerSLS2}.
For SLS of carbon nanotubes (CNTs) reinforced PA12 nanocomposite powders, finite-element heat transfer simulation was carried out to calculate the temperature distribution and melt pool's dimension~\cite{Shen2018Development}, by including laser-powder interaction, solid-liquid phase transition, and temperature-dependent material properties. Fig.~\ref{figpolymerSLS1} presents the microstructure and simulated temperature distribution after the scanning of three tracks during SLS CNTs/PA12 composite. The un-melted regions are observed in every layer due to the lack of fusion in the powder layer. The simulated depth of fusion zone is around 82 $\mu$m that is close to the experimental result.
Similar work is also reported for simulating SLS of polyamide/carbon fiber (PA/CF) and PA/NaCl composite powders by finite-element heat transfer model that incorporates phase transition heat and volumetric heat source~\cite{Tian2018Process}.
It should be noted that stereolithography (SLA) is another laser-based additive manufacturing technologies for polymers. During SLA of polymers, curing is another important difference when compared to SLS of metals and ceramics ~\cite{Hossain2015Continuum,Hossain2009Asmall,
Park20223Dprinting,Hossain2010Afinite,Hossain2009Afinite}.
In the phenomenological continuum approach, the whole curing phenomenon is used in the simulation of SLA polymer, i.e., all reactions during polymer curing is described by a set of differential equations. Recently, Brighenti et al.~\cite{Brighenti2021Laser-based} have summarized different models that could be utilized to describe the evolution of cure degree during SLA of polymer.

\section{Summary and outlook} \label{sec3}

In summary, the recent development on the modelling and simulation methodologies of sintering process across various scales including atomistic, microscopic, and macroscopic scales are overviewed.
Atomic-scale MD simulations are helpful in deciphering the atomistic sintering mechanism, but suffer from size limitations and are far away from the real sintering case. For MD simulations of sintering high-temperature ceramics, suitable potentials remain to be explored.
The microstructure-scale methods including DEM, Monte--Carlo method, and phase-field model have advantages in the calculations of microstructure evolution without the resolution of atomic movement during the sintering process. There also exist some common limitations in these microstructure models.
For instance, necessary idealizations or simplifications are required to construct these models. Due to the intrinsic length scale, with the currently available computation capability at hand, these models are still difficult to be numerically run for the real sintering compacts with a size around or above millimetre. The model parameters such as diffusion coefficients could has huge uncertainty, maybe differing by many orders of magnitude for the same material with different impurities or lattice defects. The correlation of the experimentally measurable quantities with the microscale model parameters requires delicate treatments.  Nevertheless, the latest progress on the application of DEM and phase-field model to the SLS and DMLS based additive manufacturing shows the prominent vitality of these microstructure models in additive manufacturing. The macroscopic continuum models including sintering kinetics and isotropic/anisotropic constitutive laws are beneficial for the design of sintering processing parameters to obtain the sintered parts with a desired density, porosity or shape. They are efficient in designing sintering tasks in terms of experimental and computational cost, but suffer from the absence of microstructural details and the strong dependence of model parameters on experimental data. To this end, no individual methodology could cover all the scales of sintering process and thus sequential or concurrent multiscale simulations are desirable for the across-scale understanding, prediction, and design of sintering techniques.

As an outlook, the following directions could be possibly emphasised in the next few years.

$\bullet$
Sensibly integrating the different models spanning a large range of scales to establish an ICME approach for sintering simulations remains to be explored. Thereby not only the collection of various methods is meaningful, but also well-thought scale bridging strategies should be regarded, including sequential strategy like parameter sharing and data-driven approach. One of the key issues of ICME for sintering simulation has to be resolved is the systematic integration of uncertainty in both the models and computational tools~\cite{Raether2019Simulation}, as well as the model parameters transfer across scales.

$\bullet$
New numerical techniques have to be continuously developed beyond the currently available DEM, finite element method, Monte--Carlo method, etc. Recently, material point method is applied to the particle-scale simulations of SLS~\cite{Maeshima2021Particle}. Machine-learning approaches using a multi-layered neural network, supervised machine learning techniques, linear regression models, etc. are promising for sintering simulations~\cite{Shigaki1999Amachine,Abreu2021Evaluation, Mallick2021Application,Tang2021Machine,Kim2022Tool}.

$\bullet$
Unconventional sintering technology such as microwave sintering, flash sintering, and spark plasma sintering~\cite{RENAUX20216617Mechanical, Yu2017Review, Katz1992Microwave,Guillon2014Field, Biesuz2019Flash, SERRAZINA20201205Modelling} requires modifications or redevelopments of the current models due to the involved extreme conditions, non-equilibrium states, and nonlinear multiphysics couplings. Lately, a thermo-electro-mechanical modeling of spark plasma sintering processes is proposed to incorporate the plastic and creep strain in the solid grain, surface diffusion, and grain boundary diffusion~\cite{Semenov2021Thermo}.

$\bullet$
The external-field assisted sintering technique that integrates sintering with magnetic field, electric field/current, and acoustic field~\cite{Guillon2018Manipulation, Hu2021Recent} have provided more degrees of freedom for the design of sintering process. It creates new opportunities and challenges for the multiphysics modelling of sintering.

$\bullet$
The modelling and simulation of the emerged sintering based additive manufacturing is still under development~\cite{Papazoglou2021} and a great deal of issues regarding multiphysics model and efficient numerics remain to be resolved.

\section*{CRediT authorship contribution statement}
\textbf{Min Yi:} Conceptualization, Resources, Supervision, Project administration, Funding acquisition, Writing -- original draft \& review \& editing. \textbf{Wenxuan Wang:} Conceptualization, Investigation, Data curation, Writing -- original draft. \textbf{Ming Xue:} Conceptualization, Investigation, Data curation, Writing -- original draft. \textbf{Qihua Gong:} Conceptualization, Resources, Supervision, Project administration, Writing -- original draft \& review \& editing.  \textbf{Bai-Xiang Xu:} Conceptualization, Project administration, Funding acquisition, Writing -- review.

\section*{Declaration of Competing Interest}
The authors declare that they have no known competing financial interests or personal relationships that could have appeared to influence the work reported in this paper.

\section*{Acknowledgements}
The authors acknowledge the support from 15$^\text{th}$ Thousand Youth Talents Program of China, National Science and Technology Major Project (J2019-IV-0014-0082), 
Research Fund of State Key Laboratory of Mechanics and Control of Mechanical Structures (MCMS-I-0419G01), and a project Funded by the Priority Academic Program Development of Jiangsu Higher Education Institutions. Xu would acknowledge the funding of German Science Foundation in the framework of SFB TRR270 and SFB TRR361 (Project Number 405553726 and 492661287).


\bibliographystyle{elsarticle-num}

\bibliography{sample}
   

\end{document}